\documentclass[11pt]{article}
\usepackage[margin=1in,dvips]{geometry}

\usepackage{amssymb}
\usepackage{rotating}
\usepackage{graphicx}
\usepackage[dvipsnames,usenames]{color}
\usepackage{amsmath}
\usepackage{url}
\usepackage{amsfonts}
\usepackage{paralist}
\usepackage{dialogue}
\usepackage{listings}
\usepackage{textcomp}
\usepackage{arydshln}
\usepackage{hyperref}
\usepackage{multirow}
\usepackage[font={small}]{caption, subfig}

\newcommand{\Mi}{Mixed-initiative}
\newcommand{\mi}{mixed-initiative}
\newcommand{\mii}{mixed-initiative interation}
\newcommand{\mix}{[\![\mathtt{mix}]\!]}

\begin{document}
\sloppy

\title{\bfseries Specifying and Staging Mixed-Initiative Dialogs\\with Program Generation and Transformation}
\author{\Large \bfseries Saverio Perugini\\
\small Department of Computer Science\\
\small University of Dayton\\
\small 300 College Park\\
\small Dayton, Ohio\ \ 45469--2160\ \ USA\\
\small Tel: +001 (937) 229--4079, Fax: +001 (937) 229--2193\\
\small E-mail: \url{saverio@udayton.edu}\\
\small WWW: \url{http://academic.udayton.edu/SaverioPerugini}}

\maketitle

\thispagestyle{empty}

\begin{abstract} Specifying and implementing flexible human-computer dialogs,
such as those used in kiosks and smart phone apps, is challenging because of
the numerous and varied directions in which each user might steer a dialog. The
objective of this research is to improve dialog specification and
implementation.  To do so we enriched a notation based on concepts from
programming languages, especially partial evaluation, for specifying a variety
of unsolicited reporting, \mi\ dialogs in a concise representation that serves
as a design for dialog implementation.  We also built a dialog mining system
that extracts a specification in this notation from requirements.  To
demonstrate that such a specification provides a design for dialog
implementation, we built a system that automatically generates an
implementation of the dialog, called a \textit{stager}, from it.  These two
components constitute a dialog modeling toolkit that automates dialog
specification and implementation.  These results provide a proof of concept and
demonstrate the study of dialog specification and implementation from a
programming languages perspective.  The ubiquity of dialogs in domains such as
travel, education, and health care combined with the demand for smart phone
apps provide a landscape for further investigation of these results.
\end{abstract}

\mi\ dialogs,
\mi\ interaction,
partial evaluation,
program generation,
program specialization,
program transformation,
Scheme.

\newpage

\section{Introduction}

From interactive teller machines (\textsc{itm}s),
airport and train kiosks, and smart
phone apps to installation wizards and
intelligent tutoring or training,
human-computer dialogs\footnote{A \textit{dialog} in this context refers to any
series of interactions between a user and a computer system, not necessarily
through a verbal modality. For instance, a user completing an online mortgage
application participates in a human-computer dialog.} are woven into the fabric
of our daily interactions with computer systems.  While supporting flexibility
in dialog is essential to deliver a personalized
experience to the user, it makes
the implementation challenging due to the numerous 
and varied directions in which a user
might desire to steer a dialog, all of which must be captured in an
implementation.  This problem is difficult since dialogs range in complexity
from those modeled after a simple, fixed, predefined series of questions and answers
to those that give the user a great deal of
control over the direction in which to steer the dialog.  In this article, we
discuss a model, based on concepts from programming languages,
especially partial evaluation, for specifying and staging\footnote{We are
not referring to \textit{staging} as the `language construct that
allows a program at one stage of evaluation to manipulate and specialize
a program to be executed at a later stage'~\cite{stagingMLHOSC}.
Rather we are using the term to refer to
`staging the progressive interaction of a human-computer dialog.'} dialogs.
The objective of this research is to improve dialog specification and
implementation, and to enrich and demonstrate the feasibility of an alternate
method of modeling human-computer dialogs.

This article is organized as follows. To introduce the reader to the wide range
of dialogs possible, and provide a better feel for this problem and its
difficulty, we first present some illustrative examples of dialogs in
Section~\ref{sec:dialogs}.  We simply showcase a variety of dialogs and their
specification using formal notation, rather than discuss their implementation
and related issues which are covered later. In Section~\ref{sec:notation},
we describe a notation, based on concepts from programming languages, for
specifying dialogs.  Section~\ref{sec:staging} demonstrates how dialogs can be
staged with partial evaluation, while Section~\ref{sec:implementation} outlines
how to mine dialog specifications in this notation from dialog requirements,
and how we automatically generate stagers from those specifications.
Section~\ref{sec:discussion} summarizes our contributions and discusses future
work.

\section{Dialogs}
\label{sec:dialogs}

Let us consider a variety of dialogs we might want to model.

\subsection{Fixed- and Mixed-initiative Dialogs}

Consider a dialog to purchase gasoline using a credit card.  The customer must
first swipe the card, then choose a grade of octane, and finally indicate
whether she desires a receipt.  Such a dialog is a \textit{fixed} dialog due to
the fixed order of the questions from which the user is not permitted to
deviate in her responses~\cite{MII-UR}.

An \textit{enumerated specification} is a set of episodes, and an
\textit{episode} is an ordered list of questions to be posed and answered from
the start of the dialog through dialog completion.  Intuitively, a
specification is a complete set of all possible ways to complete a dialog.
Formally, a dialog specification is a set of \textit{totally ordered sets}.  We
use a \textit{Hasse} diagram, a graphical
depiction of a \textit{partially ordered set}, to represent a dialog
specification.  A relation $R$ with the set $S$ over whose Cartesian product
$R$ is defined is a \textit{strict
partially ordered set} (or \textit{poset}) if $R$
is an irreflexive, asymmetric, and transitive relation.  This means that some
of the elements of $S$ may be unordered based on the relation $R$.  On the
other hand,
a set $S$ is a \textit{strict
totally ordered set} according to a relation $R$ if and only if for
every two elements $(x,y) \in S$, $x R y$ or $y R x$.  Every totally ordered
set is also a partially ordered set, but the reverse is not necessarily true.

\begin{table}
\centering
\begin{minipage}{\linewidth}
\caption{A spectrum of dialogs
from fixed (a, second column) to complete, \mi\ dialogs (e, sixth column),
encompassing a variety of unsolicited reporting, \mi\ dialogs, in
three representations: enumerated specification (second row), Hasse diagam (third row), and
our notation using concepts from programming languages (fourth row).
Last (fifth) row gives staging expression, using partial evaluation ($\mix$),
used to stage each dialog.\protect\footnote{\tiny
The use of ellipses in the expressions in the last row (e.g., \texttt{size=\dots}) indicate that
the value of the parameter can be any valid response.}}
\label{tab:posets}
\end{minipage}
\setlength{\tabcolsep}{.001em}
\resizebox{\textwidth}{!}{
\begin{tabular}{|l|l|l|l|l|l|}

\multicolumn{1}{l}{} &
\multicolumn{1}{l}{} &
\multicolumn{1}{l}{} &
\multicolumn{1}{l}{} &
\multicolumn{1}{l}{} &
\multicolumn{1}{l|}{\scriptsize \textbf{complete,}}\\

\multicolumn{1}{l}{} &
\multicolumn{1}{|l}{\scriptsize $\longleftarrow$
\textbf{(most rigid) fixed dialogs \dots\dots\dots\dots}} &
\multicolumn{1}{c}{\scriptsize \textbf{\dots\dots\dots\dots\dots\dots\dots\dots\dots}} &
\multicolumn{1}{c}{\scriptsize \textbf{\dots\dots\dots\dots\dots\dots\dots\dots\dots\dots\dots\dots}} &
\multicolumn{1}{c}{\scriptsize \textbf{\dots\dots\dots\dots\dots\dots\dots\dots}} &
\multicolumn{1}{r|}{\scriptsize \textbf{\mi\ dialogs (most flexible) $\longrightarrow$}}\\

\hline

\multirow{1}{*}{\begin{sideways}\parbox[c]{4mm}{\begin{center}{\tiny \textbf{ID}}\end{center}}\end{sideways}} &

\multicolumn{1}{c|}{\tiny \textbf{(a)}} &
\multicolumn{1}{c|}{\tiny \textbf{(b)}} &
\multicolumn{1}{c|}{\tiny \textbf{(c)}} &
\multicolumn{1}{c|}{\tiny \textbf{(d)}} &
\multicolumn{1}{c|}{\tiny \textbf{(e)}}\\

\hline

\multirow{1}{*}{\begin{sideways}\parbox[c]{17mm}{\begin{center}{\tiny \textbf{Enumerated Specification}}\end{center}}\end{sideways}} &

\multicolumn{1}{c|}{\tiny
\begin{tabular}{l}
\{$\prec$\textrm{credit-card grade receipt}$\succ$\}
\end{tabular}} &

\multicolumn{1}{c|}{\tiny
\begin{tabular}{l}
\{$\prec$\textsc{pin} transaction account amount$\succ$,\\
\ \,$\prec$\textsc{pin} account transaction amount$\succ$\}
\end{tabular}} &

\multicolumn{1}{c|}{\tiny
\begin{tabular}{l}
\{$\prec$receipt sandwich beverage dine-in/take-out$\succ$,\\
\ \,$\prec$dine-in/take-out sandwich beverage receipt$\succ$\}
\end{tabular}} &

\multicolumn{1}{c|}{\tiny
\begin{tabular}{l}
\{$\prec$cream sugar eggs toast$\succ$,\\
\ \,$\prec$cream sugar toast eggs$\succ$,\\
\ \,$\prec$(cream sugar) toast eggs$\succ$,\\
\ \,$\prec$(cream sugar) eggs toast$\succ$,\\
\ \,$\prec$sugar cream eggs toast$\succ$,\\
\ \,$\prec$sugar cream toast eggs$\succ$,\\
\ \,$\prec$eggs toast cream sugar$\succ$,\\
\ \,$\prec$eggs toast sugar cream$\succ$,\\
\ \,$\prec$toast eggs cream sugar$\succ$,\\
\ \,$\prec$toast eggs sugar cream$\succ$,\\
\ \,$\prec$sugar cream (eggs toast)$\succ$,\\
\ \,$\prec$cream sugar (eggs toast)$\succ$,\\
\ \,$\prec$(eggs toast) (cream sugar)$\succ$,\\
\ \,$\prec$(cream sugar) (eggs toast)$\succ$\}
\end{tabular}} &

\multicolumn{1}{c|}{\tiny
\begin{tabular}{l}
\{$\prec$(size blend cream)$\succ$,\\
\ \,$\prec$(size blend) cream$\succ$,\\
\ \,$\prec$cream (size blend)$\succ$,\\
\ \,$\prec$(blend cream) size$\succ$,\\
\ \,$\prec$size (blend cream)$\succ$,\\
\ \,$\prec$(size cream) blend$\succ$,\\
\ \,$\prec$blend (size cream)$\succ$,\\
\ \,$\prec$size blend cream$\succ$,\\
\ \,$\prec$size cream blend$\succ$,\\
\ \,$\prec$blend size cream$\succ$,\\
\ \,$\prec$blend cream size$\succ$,\\
\ \,$\prec$cream blend size$\succ$,\\ 
\ \,$\prec$cream size blend$\succ$\}
\end{tabular}}\\

\hline

\multirow{1}[1]{*}{\begin{sideways}\parbox[t]{16mm}{\begin{center}{\tiny \textbf{Hasse diagram}}\end{center}}\end{sideways}} &

\multicolumn{1}{c|}{\includegraphics[scale=0.42]{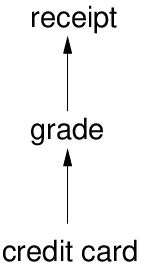}} &
\multicolumn{1}{c|}{\includegraphics[scale=0.42]{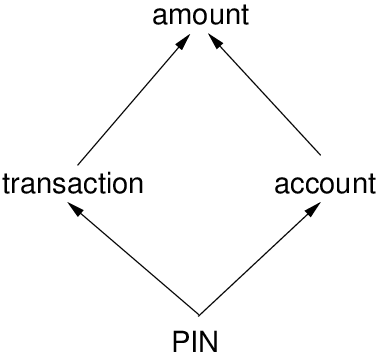}} &
\multicolumn{1}{c|}{\includegraphics[scale=0.42]{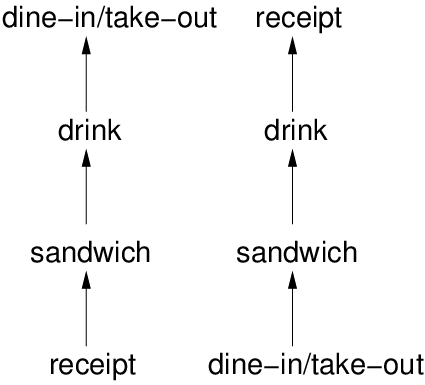}} &
\multicolumn{1}{c|}{\includegraphics[scale=0.42]{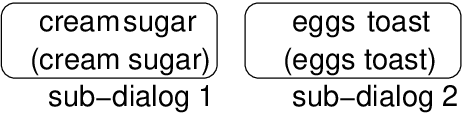}} &
\multicolumn{1}{c|}{\includegraphics[scale=0.42]{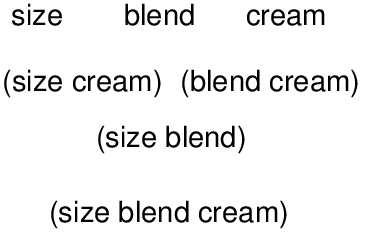}}\\

\hline

\multirow{1}[1]{*}{\begin{sideways}\parbox[c]{15mm}{\begin{center}{\tiny \textbf{PL Notation}}\end{center}}\end{sideways}} &

\multicolumn{1}{c|}{\scriptsize $\frac{C}{\textrm{credit-card grade receipt}}$} &

\multicolumn{1}{c|}{\scriptsize $\frac{C}{\textrm{PIN} \frac{SPE^{'}}{\textrm{transaction account}} \textrm{amount}}$} &

\multicolumn{1}{c|}{\scriptsize
\begin{tabular}{l}
\multicolumn{1}{c}{$\frac{C}{\textrm{receipt sandwich drink dine-in/take-out}}$}\\
\multicolumn{1}{c}{$\cup$}\\
\multicolumn{1}{c}{\scriptsize $\frac{C}{\textrm{dine-in/take-out sandwich drink receipt}}$}
\end{tabular}} &

\multicolumn{1}{c|}{\scriptsize $\frac{SPE^{'}}{\frac{{PE}^{\star}}{\textrm{cream sugar}} \frac{{PE}^{\star}}{\textrm{eggs toast}}}$} &

\multicolumn{1}{c|}{\scriptsize $\frac{{PE}^{\star}}{\textrm{size blend cream}}$}\\

\hline

\multirow{1}[1]{*}{\begin{sideways}\parbox[c]{15mm}{\begin{center}{\tiny \textbf{Implementation}}\end{center}}\end{sideways}} &

\multicolumn{1}{c|}{\tiny
\begin{tabular}{l}
$\mix [\mix [\mix [\mathrm{f}, \mathrm{size=\dots}], \mathrm{blend=\dots}], \mathrm{cream=\dots}]$ 
\end{tabular}} & & & &

\multicolumn{1}{c|}{\tiny
\begin{tabular}{l}
$\mix [\mathrm{f}, \mathrm{size, blend, cream}]$,\\
$\mix [\mix [\mathrm{f}, \mathrm{size=\dots}], \mathrm{blend=\dots, cream=\dots}]$,\\
$\mix [\mix [\mathrm{f}, \mathrm{blend=\dots, cream=\dots}], \mathrm{size=\dots}]$,\\
$\mix [\mix [\mathrm{f}, \mathrm{blend=\dots}], \mathrm{cream=\dots, size=\dots}]$,\\
$\mix [\mix [\mathrm{f}, \mathrm{cream=\dots, size=\dots}], \mathrm{blend=\dots}]$,\\
$\mix [\mix [\mathrm{f}, \mathrm{cream=\dots}], \mathrm{size=\dots, blend=\dots}]$,\\
$\mix [\mix [\mathrm{f}, \mathrm{size=\dots, blend=\dots}], \mathrm{cream=\dots}]$,\\
$\mix [\mix [\mix [\mathrm{f}, \mathrm{size=\dots}], \mathrm{cream=\dots}], \mathrm{blend=\dots}]$,\\
$\mix [\mix [\mix [\mathrm{f}, \mathrm{size=\dots}], \mathrm{blend=\dots}], \mathrm{cream=\dots}]$,\\
$\mix [\mix [\mix [\mathrm{f}, \mathrm{blend=\dots}], \mathrm{size=\dots}], \mathrm{cream=\dots}]$,\\
$\mix [\mix [\mix [\mathrm{f}, \mathrm{blend=\dots}], \mathrm{cream=\dots}], \mathrm{size=\dots}]$,\\
$\mix [\mix [\mix [\mathrm{f}, \mathrm{cream=\dots}], \mathrm{size=\dots}], \mathrm{blend=\dots}]$,\\
$\mix [\mix [\mix [\mathrm{f}, \mathrm{cream=\dots}], \mathrm{blend=\dots}], \mathrm{size=\dots}]$
\end{tabular}}\\

\hline
\end{tabular}}
\end{table}

An enumerated
specification of this gasoline dialog is 
{\scriptsize 
\{$\prec$\textrm{credit-card grade receipt}$\succ$\}}, and
Table~\ref{tab:posets}a illustrates the Hasse diagram that
specifies it.
A Hasse diagram is read bottom-up.  Here, the set $S$ of the
poset is the set of the questions posed in the dialog and $R$ of the poset is
the `must be answered before'
relation denoted with an upward arrow between the source and target of the
arrow.

Flexible dialogs typically support multiple completion paths.  For instance,
consider a dialog for ordering coffee.  The participant must select a
size and blend, and indicate whether room for cream is desired.
Since possible responses to these questions
are completely independent of each other, the
dialog designer may wish to permit the
participant to communicate the answers in any combinations and in any order.
For example, some customers may prefer to use a
{\scriptsize $\prec$\textrm{size blend cream}$\succ$} episode:

{\scriptsize
\begin{dialogue}
\speak{System} Please select a size.
\speak{User} Small.
\speak{System} Please select a blend.
\speak{User} Dark.
\speak{System} Please specify whether you want room for cream.
\speak{User} No room for cream.
\end{dialogue}}

\noindent
Others may prefer a
{\scriptsize $\prec$\textrm{blend cream size}$\succ$} episode:

{\scriptsize
\begin{dialogue}
\speak{System} Please select a blend.
\speak{User} Mild.
\speak{System} Please specify whether you want room for cream.
\speak{User} With room for cream.
\speak{System} Please select a size.
\speak{User} Large.
\end{dialogue}}

\noindent
Still others might prefer to use a
{\scriptsize $\prec$\textrm{(size blend) cream}$\succ$} episode,
where answers to the questions enclosed in parentheses 
must be communicated in a single utterance (i.e., all at once):

{\scriptsize
\begin{dialogue}
\speak{System} Please select a size.
\speak{User} Small, french roast.
\speak{System} Please specify whether you want room for cream.
\speak{User} No.
\end{dialogue}}

\noindent
Therefore, to accommodate all possibilities we specify this dialog as:

\begin{center}
{\scriptsize
\begin{tabular}{lll}
\{$\prec$(size blend cream)$\succ$, &
$\prec$(size blend) cream$\succ$, &
$\prec$cream (size blend)$\succ$,\\
\ \,$\prec$(blend cream) size$\succ$, &
$\prec$size (blend cream)$\succ$, &
$\prec$(size cream) blend$\succ$,\\
\ \,$\prec$blend (size cream)$\succ$, &
$\prec$size blend cream$\succ$, &
$\prec$size cream blend$\succ$,\\
\ \,$\prec$blend size cream$\succ$, &
$\prec$blend cream size$\succ$, &
$\prec$cream blend size$\succ$,\\ 
\ \,$\prec$cream size blend$\succ$\}.
\end{tabular}}
\end{center}

Notice that this
specification indicates that answers to the set of questions in the dialog
may be communicated in utterances corresponding to all
possible set partitions of the set of questions.  Moreover, all possible
permutations of those partitions are specified as well.  The Hasse diagram for
this dialog is given in Table~\ref{tab:posets}e.
The absence of arrows between the size, blend, cream,
(size blend), (size cream), (blend cream), and (size blend cream) elements
indicates that the times at which each of those utterances may
be communicated are unordered.  Notice that the Hasse
diagram is a compressed representation capturing the
requirements in the specification.  Moreover, the compression is lossless
(i.e., the episodes in the enumerated
specification may be reconstructed from the diagram).

Giving the user more flexibility in how to proceed through a
dialog increases the number of episodes in its enumerated specification. 
This coffee ordering dialog  
is a \textit{\mi} dialog~\cite{MII-UR}. There are multiple
tiers of \mi\ interaction.  The tier considered in this article is called \textit{
unsolicited reporting}---an interaction strategy where, in
response to a question, at any point in the dialog, the user may provide an
unsolicited response to a forthcoming question.

\begin{table}
\centering
\caption{Sample dialogs involving permutations or
partitions of responses to questions.
Map associating a concept from programming
languages to each element of the cross-product of utterance orders 
in leftmost column and utterance sizes in topmost row. Bolded parenthesized
letters connect these dialogs to those in Table~\ref{tab:posets}.}
{\scriptsize
\begin{tabular}{lll}
\hline\noalign{\smallskip}
& \textbf{Only a single response} & \textbf{Multiple responses}\\
& \textbf{per utterance} & \textbf{per utterance}\\
\noalign{\smallskip}\hline\noalign{\smallskip}
\textbf{Only one} & Confirmation dialog boxes & Online forms with\\
\textbf{utterance} & common in application software; & multiple fields;\\
& interpretation ($I$) & interpretation ($I$)\\
\noalign{\smallskip}\hline\noalign{\smallskip}
\textbf{Totally} & Purchasing gasoline with a & Providing a telephone,\\
\textbf{ordered} & credit card; buying beverages &
credit card, or \textsc{pin} number\\
\textbf{utterances} & from a vending machine; & through voice;\\
& currying ($C$) \textbf{(a)} & partial function\\
& & application $n$ ($PFA_{n}^{\star}$)\\
\noalign{\smallskip}\hline\noalign{\smallskip}
\textbf{Partially} &
\textsc{itm}s, and airport or train kiosks; &
Ordering a coffee or pizza;\\
\textbf{ordered} & single-argument & partial evaluation ($PE^{\star}$) \textbf{(e)}\\
\textbf{utterances} & partial evaluation ($SPE^{'}$) \textbf{(b)} &\\
\noalign{\smallskip}\hline
\end{tabular}}
\label{tab:pracspace}
\end{table}

When all possible permutations (i.e., orders) of all possible
partitions (i.e., combinations) of responses to questions are supported, we
call the dialog a \textit{complete}, \mi\ dialog.  Table~\ref{tab:posets}
represents a space from fixed to complete,
\mi\ dialogs, encompassing a wide variety of
unsolicited reporting, \mi\ dialogs. Table~\ref{tab:pracspace}
identifies some practical, everyday dialogs that fall into
the cross product of permutations and partitions of
responses to questions.

\begin{figure}
\centering
\includegraphics[scale=0.50]{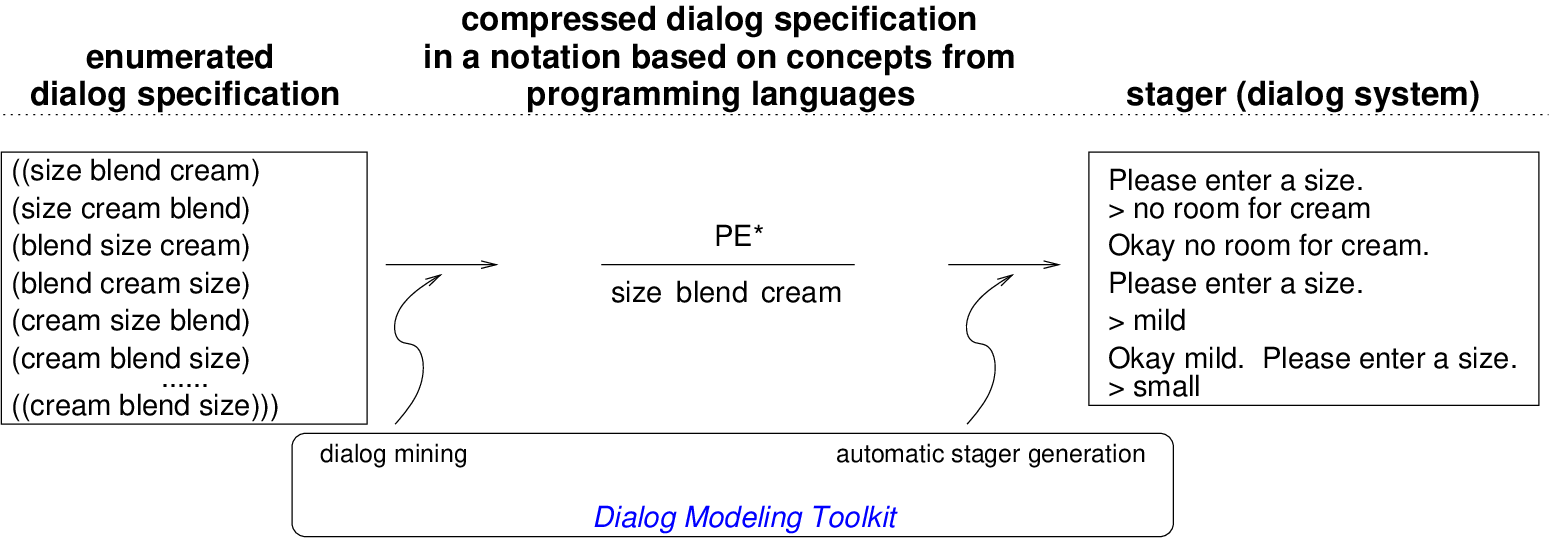}
\caption{Conceptual overview of our research project through an example.}
\label{fig:overview}
\end{figure}

Fig.~\ref{fig:overview} provides an overview of this research project.  We start
with an enumerated dialog specification (i.e., a set of episodes) and mine it
for a compressed representation of the dialog in a notation based on concepts
from programming languages that captures the requirements of the dialog
(transition from the left to the center of Fig.~\ref{fig:overview})---a process we
call \textit{dialog mining}.
From that intermediate, implementation-neutral representation we automatically
generate a \textit{stager}---a program capable of realizing and executing the
dialog or, in other words, `staging' the progressive interaction' (transition
from the center to the right of Fig.~\ref{fig:overview}).

\subsection{Spectrum of Dialogs}

Fixed and complete, \mi\ dialogs represent opposite ends of this spectrum of
dialogs shown in Table~\ref{tab:posets} that encompass several dialogs.  For
instance, consider a specification for an \textsc{itm} dialog where
\textsc{pin} and amount must be entered first and last, respectively, but the
transaction type (e.g., deposit or withdrawal) and account type (e.g., checking
or savings) may be communicated in any order (see Table~\ref{tab:posets}b):

\begin{center}
{\scriptsize
\{$\prec$\textsc{pin} transaction account amount$\succ$, 
$\prec$\textsc{pin} account transaction amount$\succ$\}}.
\end{center}

\noindent This dialog contains an embedded, \mi\ sub-dialog~\cite{MII-UR}:
\{$\prec$transaction account$\succ$, 
$\prec$account transaction$\succ$\}.

Alternatively, consider a dialog for ordering lunch where requesting a receipt
or indicating whether you are dining-in or taking-out can be communicated
either first or last, but specification of sandwich and beverage must occur in
that order:

\begin{center}
{\scriptsize
\begin{tabular}{l}
\{$\prec$receipt sandwich beverage dine-in/take-out$\succ$,
$\prec$dine-in/take-out sandwich beverage receipt$\succ$\}.
\end{tabular}}
\end{center}

\noindent This dialog contains an embedded, fixed sub-dialog (i.e.,
{\scriptsize 
\{$\prec$sandwich beverage$\succ$\}}) and, unlike the prior examples, cannot be
captured by a single poset (see Table~\ref{tab:posets}c).

Lastly, consider a dialog containing two embedded,
complete, \mi\ sub-dialogs~\cite{MIIMC} (see Table~\ref{tab:posets}d):

\begin{center}
{\scriptsize
\begin{tabular}{lll}
\{$\prec$cream sugar eggs toast$\succ$, &
$\prec$cream sugar toast eggs$\succ$, &
$\prec$(cream sugar) toast eggs$\succ$,\\
\ \,$\prec$(cream sugar) eggs toast$\succ$, &
$\prec$sugar cream eggs toast$\succ$, &
$\prec$sugar cream toast eggs$\succ$,\\
\ \,$\prec$eggs toast cream sugar$\succ$, &
$\prec$eggs toast sugar cream$\succ$, &
$\prec$toast eggs cream sugar$\succ$,\\
\ \,$\prec$toast eggs sugar cream$\succ$, &
$\prec$sugar cream (eggs toast)$\succ$, &
$\prec$cream sugar (eggs toast)$\succ$,\\
\ \,$\prec$(eggs toast) (cream sugar)$\succ$, &
$\prec$(cream sugar) (eggs toast)$\succ$\}.
\end{tabular}}
\end{center}

\noindent Here, the user can specify coffee and breakfast choices in any order,
and can specify the sub-parts of coffee and breakfast in any order, but cannot
mix the atomic responses of the two (i.e., episodes such as
{\scriptsize $\prec$cream eggs sugar toast$\succ$} are not permitted).

There are two assumptions we make in this article on this spectrum of
dialogs: i) each episode in a specification has 
a consistent number of questions and ii) the permissible responses
for each question are completely independent of each other (i.e., no response
to a question ever precludes a particular response to another question).

\section{Specifying \Mi\ Dialogs with Concepts of Languages}
\label{sec:notation}

There is a combinatorial explosion in the number of possible dialogs between
the fixed and complete, \mi\ ends of the spectrum in Table~\ref{tab:posets}.
Specifically,
the number of dialogs possible in this space is
{\scriptsize $2^{|\mathcal{D}_{cmi}|}-1 = \sum_{r=1}^{|\mathcal{D}_{cmi}|}
{{|\mathcal{D}_{cmi}|}\choose{r}}$}
(i.e., all possible subsets, save for the empty
set, of all episodes in a complete, \mi\ dialog),
where
{\scriptsize $\mathcal{D}_{cmi}$}
represents the enumerated specification of a complete,
\mi\ dialog given $q$, the number of questions posed in the dialog.
In this section we bring structure to this
space by viewing these dialogs
through a programming languages lens.
This approach also lends insight
into staging them. We start by describing how to specify these dialogs using a
notation based on a variety of concepts from programming languages.

\subsection{Notation}

\begin{table}
\centering
\caption{Specifications of dialogs in a notation
based on concepts from programming languages (second row) and
as enumerated specifications (third row).
Last (fourth) row gives staging expression, using partial evaluation ($\mix$),
used to stage each dialog.}
\setlength{\tabcolsep}{.001em}
\resizebox{\textwidth}{!}{
\begin{tabular}{|l|l|l|l|l|l|}

\multicolumn{1}{l}{} &
\multicolumn{5}{|c|}{\scriptsize $\longleftarrow$ \textbf{\dots dialogs
between fixed dialogs and complete, \mi\ dialogs ($\Delta$) \dots} $\longrightarrow$}\\

\hline

\multirow{1}{*}{\begin{sideways}\parbox[c]{4mm}{\begin{center}{\tiny \textbf{ID}}\end{center}}\end{sideways}} &

\multicolumn{1}{c|}{\tiny \textbf{(f)}} &
\multicolumn{1}{c|}{\tiny \textbf{(g)}} &
\multicolumn{1}{c|}{\tiny \textbf{(h)}} &
\multicolumn{1}{c|}{\tiny \textbf{(i)}} &
\multicolumn{1}{c|}{\tiny \textbf{(j)}}\\

\hline

\multirow{1}[1]{*}{\begin{sideways}\parbox[c]{8mm}{\begin{center}{\tiny \textbf{PL Not.}}\end{center}}\end{sideways}} &

\multicolumn{1}{c|}{\scriptsize $\frac{PFA_n}{\textrm{size blend cream}}$} &

\multicolumn{1}{c|}{\scriptsize $\frac{{PFA}_{n}^{\star}}{\textrm{size blend cream}}$} &

\multicolumn{1}{c|}{\scriptsize $\frac{SPE}{\textrm{size blend cream}}$} &

\multicolumn{1}{c|}{\scriptsize $\frac{SPE^{'}}{\textrm{size blend cream}}$} &

\multicolumn{1}{c|}{\scriptsize $\frac{PE}{\textrm{size blend cream}}$}\\

\hline

\multirow{1}{*}{\begin{sideways}\parbox[c]{10mm}{\begin{center}{\tiny \textbf{Enum. Spec.}}\end{center}}\end{sideways}} &

\multicolumn{1}{c|}{\tiny
\begin{tabular}{l}
\{$\prec$\textrm{(size blend cream)}$\succ$,\\
\ \,$\prec$\textrm{(size (blend cream)}$\succ$,\\
\ \,$\prec$\textrm{(size blend) cream}$\succ$\}
\end{tabular}} &

\multicolumn{1}{c|}{\tiny
\begin{tabular}{l}
\{$\prec$\textrm{(size blend cream)}$\succ$,\\
\ \,$\prec$\textrm{size (blend cream)}$\succ$,\\
\ \,$\prec$\textrm{(size blend) cream}$\succ$,\\
\ \,$\prec$\textrm{size blend cream}$\succ$\}
\end{tabular}} &

\multicolumn{1}{c|}{\tiny
\begin{tabular}{l}
\{$\prec$\textrm{size (blend cream)}$\succ$,\\
\ \,$\prec$\textrm{blend (size cream)}$\succ$,\\
\ \,$\prec$\textrm{cream (size blend)}$\succ$\}
\end{tabular}} &

\multicolumn{1}{c|}{\tiny
\begin{tabular}{l}
\{$\prec$\textrm{size blend cream}$\succ$,\\
\ \,$\prec$\textrm{size cream blend}$\succ$,\\
\ \,$\prec$\textrm{blend size cream}$\succ$,\\
\ \,$\prec$\textrm{blend cream size}$\succ$,\\
\ \,$\prec$\textrm{cream blend size}$\succ$,\\
\ \,$\prec$\textrm{cream size blend}$\succ$\}
\end{tabular}} &

\multicolumn{1}{c|}{\tiny
\begin{tabular}{l}
\{$\prec$\textrm{(size blend cream)}$\succ$,\\
\ \,$\prec$\textrm{size (blend cream)}$\succ$,\\
\ \,$\prec$\textrm{blend (size cream)}$\succ$,\\
\ \,$\prec$\textrm{cream (size blend)}$\succ$,\\
\ \,$\prec$\textrm{(size blend) cream}$\succ$,\\
\ \,$\prec$\textrm{(size cream) blend}$\succ$,\\
\ \,$\prec$\textrm{(blend cream) size}$\succ$\}
\end{tabular}}\\

\hline

\multirow{1}[1]{*}{\begin{sideways}\parbox[c]{8mm}{\begin{center}{\tiny \textbf{Impl.}}\end{center}}\end{sideways}} &

\multicolumn{1}{c|}{\tiny
\begin{tabular}{l}
$\mix [\mathrm{f}, \mathrm{size=\dots, blend=\dots, cream=\dots}]$,\\
$\mix [\mix [\mathrm{f}, \mathrm{size=\dots}], \mathrm{blend=\dots, cream=\dots}]$,\\
$\mix [\mix [\mathrm{f}, \mathrm{size=\dots, blend=\dots}], \mathrm{cream=\dots}]$
\end{tabular}} & 

\multicolumn{1}{c|}{\tiny
\begin{tabular}{l}
$\mix [\mathrm{f}, \mathrm{size=\dots, blend=\dots, cream=\dots}]$,\\
$\mix [\mix [\mathrm{f}, \mathrm{size=\dots}], \mathrm{blend=\dots, cream=\dots}]$,\\
$\mix [\mix [\mathrm{f}, \mathrm{size=\dots, blend=\dots}], \mathrm{cream=\dots}]$,\\
$\mix [\mix [\mix [\mathrm{f}, \mathrm{size=\dots}], \mathrm{blend=\dots}], \mathrm{cream=\dots}]$
\end{tabular}} & 

\multicolumn{1}{c|}{\tiny
\begin{tabular}{l}
$\mix [\mix [\mathrm{f}, \mathrm{size=\dots}], \mathrm{cream=\dots, blend=\dots}]$,\\
$\mix [\mix [\mathrm{f}, \mathrm{blend=\dots}], \mathrm{size=\dots, cream=\dots}]$,\\
$\mix [\mix [\mathrm{f}, \mathrm{cream=\dots}], \mathrm{size=\dots, blend=\dots}]$
\end{tabular}} &

\multicolumn{1}{c|}{\tiny
\begin{tabular}{l}
$\mix [\mix [\mix [\mathrm{f}, \mathrm{size=\dots}], \mathrm{cream=\dots}], \mathrm{blend=\dots}]$,\\
$\mix [\mix [\mix [\mathrm{f}, \mathrm{size=\dots}], \mathrm{blend=\dots}], \mathrm{cream=\dots}]$,\\
$\mix [\mix [\mix [\mathrm{f}, \mathrm{blend=\dots}], \mathrm{size=\dots}], \mathrm{cream=\dots}]$,\\
$\mix [\mix [\mix [\mathrm{f}, \mathrm{blend=\dots}], \mathrm{cream=\dots}], \mathrm{size=\dots}]$,\\
$\mix [\mix [\mix [\mathrm{f}, \mathrm{cream=\dots}], \mathrm{size=\dots}], \mathrm{blend=\dots}]$,\\
$\mix [\mix [\mix [\mathrm{f}, \mathrm{cream=\dots}], \mathrm{blend=\dots}], \mathrm{size=\dots}]$
\end{tabular}} &

\multicolumn{1}{c|}{\tiny
\begin{tabular}{l}
$\mix [\mathrm{f}, \mathrm{size=\dots, blend=\dots, cream=\dots}]$,\\
$\mix [\mix [\mathrm{f}, \mathrm{size=\dots}], \mathrm{cream=\dots, blend=\dots}]$,\\
$\mix [\mix [\mathrm{f}, \mathrm{blend=\dots}], \mathrm{size=\dots, cream=\dots}]$,\\
$\mix [\mix [\mathrm{f}, \mathrm{cream=\dots}], \mathrm{size=\dots, blend=\dots}]$,\\
$\mix [\mix [\mathrm{f}, \mathrm{size=\dots, blend=\dots}], \mathrm{cream=\dots}]$,\\
$\mix [\mix [\mathrm{f}, \mathrm{size=\dots, cream=\dots}], \mathrm{blend=\dots}]$,\\
$\mix [\mix [\mathrm{f}, \mathrm{blend=\dots, cream=\dots}], \mathrm{size}]$
\end{tabular}}\\

\hline
\end{tabular}}

\label{tab:others}
\end{table}

We developed a computational model for specifying and implementing \mi\
dialogs, centered around this `model one path, yet support many paths' theme.
The fundamental aspect of our model supporting that theme is our novel use of
program transformations (e.g., partial evaluation) and other concepts from
programming languages (e.g., functional currying) to specify and stage \mi\
dialogs.

Our model involves a notation for dialog specification in a
compressed manner using these and other concepts~\cite{vt-staging,WWW2004}.
In this notation a dialog is specified by an expression of the form
$\frac{X}{T}$, where $X$ represents a program transformation or language
concept and $T$ represents a list of terms, where each term represents either a
question (of the dialog) or a sub-dialog (introduced below) in the dialog being
specified.  Each expression represents a set of episodes.
The main thematic idea is that the set of
episodes specified by an expression of this form correspond to all possible
ways that a function parameterized by the terms (e.g., dialog questions) in the
denominator can be partially applied, and re-partially applied, and so on
progressively, according to the semantics of the transformation operator or
language concept in the numerator.\footnote{This notation was introduced
in~\cite{vt-staging} and revised in~\cite{WWW2004}.  Here, we enrich it with
additional concepts and modify its semantics.}
For instance, we use the concept of $I$\textit{nterpretation}~\cite{EOPL3} to
specify a dialog where \textit{all} the responses to all dialog questions must
be communicated in a single utterance (e.g., Q: `Please select a size.'
A: `Small, dark roast, no cream.'), such as {\scriptsize
$\frac{I}{\textrm{blend cream size}}$ = \{$\prec$(blend cream
size)$\succ$\}},
because interpreting a
function requires that \textit{all} arguments be supplied at the time of the
call, corresponding to a complete evaluation.  Similarly, we use the concept
of $C$\textit{urrying} to specify a fixed dialog, such as {\scriptsize
$\frac{C}{\textrm{blend cream size}}$ = \{$\prec$blend cream size$\succ$\}},
where only one fixed episode is permitted.  Currying transforms a function
$f_{uncurried}$ with type signature $(p_1 \times p_2 \times \cdots \times p_n)
\rightarrow r$ to a function $f_{curried}$ with type signature $p_1 \rightarrow
(p_2 \rightarrow (\cdots \rightarrow (p_n \rightarrow r) \cdots)$, such that
$f_{uncurried}(a_1, a_2, \cdots, a_n) = (\cdots  ((f_{curried} (a_1)) (a_2))
\cdots) (a_n)$.  Currying $f_{uncurried}$ and running the resulting
$f_{curried}$ function has the same effect as progressively partially applying
$f_{uncurried}$, resulting in a dialog spread across multiple stages of
interaction (i.e., questions and answers), but still in a fixed, prescribed
order (e.g., Q: `Please select a size.' A: `Small,' Q: `Please select a blend.'
A: `Dark roast,' Q: `Please specify whether you want room for cream.' A: `Yes.').

\begin{table}
\centering
\caption{Type signatures of functions in the model demonstrating
that partial evaluation ($\mix$)
subsumes the other functions here. Assumes a ternary
function $f: (a \times b \times c) \rightarrow d$.}
{\scriptsize
\begin{tabular}{llllll}
\hline\noalign{\smallskip}
\multicolumn{1}{c}\textbf{Concept} & \multicolumn{1}{c}\textbf{Function} & & \multicolumn{1}{c}\textbf{Type signature} & & \multicolumn{1}{c}{$\mathbf{\lambda}$\textbf{-calculus}}\\
\noalign{\smallskip}\hline\noalign{\smallskip}
$I$ & \texttt{apply} & : & $(((a \times b \times c) \rightarrow d) \times a \times b \times c) \rightarrow d$  & = &
$\lambda(f, x, y, z). f(x, y, z)$\\

$C$ & \texttt{curry} & : &
$((a \times b \times c) \rightarrow d) \rightarrow (a \rightarrow (b \rightarrow (c \rightarrow d)))$
& = & 
$\lambda(f). \lambda(x). \lambda(y). \lambda(z). f(x, y, z)$\\

$PFA_1$ & \texttt{papply1} & : &
$(((a \times b \times c) \rightarrow d) \times a) \rightarrow ((b \times c) \rightarrow d)$
& = & 
$\lambda(f, x). \lambda (y, z). f(x, y, z)$\\

$PFA_n$ & \texttt{papplyn} & : &
$(((a \times b \times c) \rightarrow d) \times a) \rightarrow ((b \times c) \rightarrow d)$
& = & $\lambda(f, x). \lambda(y, z). f(x, y, z)$\\

& & & $(((a \times b \times c) \rightarrow d) \times a \times b) \rightarrow (c \rightarrow d)$ & = & $\lambda(f, x, y). \lambda(z). f(x, y, z)$\\

& & & $(((a \times b \times c) \rightarrow d) \times a \times b \times c) \rightarrow (\{\} \rightarrow d)$ & = & $\lambda(f, x, y, z). \lambda(). f(x, y, z)$\\

$SPE$ & \texttt{smix} & : &
$(((a \times b \times c) \rightarrow d) \times a) \rightarrow ((b \times c) \rightarrow d)$
& = & $\lambda(f,x). \lambda(y,z). f(x,y,z)$\\

& & & $(((a \times b \times c) \rightarrow d) \times b) \rightarrow ((a \times c) \rightarrow d)$ & = & $\lambda(f,y). \lambda(x,z). f(x,y,z)$\\

& & & $(((a \times b \times c) \rightarrow d) \times c) \rightarrow ((a \times b) \rightarrow d)$ & = & $\lambda(f,z). \lambda(x,y). f(x,y,z)$\\

$PE$ & \texttt{mix} & : &
$(((a \times b \times c) \rightarrow d) \times a) \rightarrow ((b \times c) \rightarrow d)$
& = & $\lambda(f,x). \lambda(y,z). f(x,y,z)$\\

& & & $(((a \times b \times c) \rightarrow d) \times b) \rightarrow ((a \times c) \rightarrow d)$ & = & $\lambda(f,y). \lambda(x,z). f(x,y,z)$\\

& & & $(((a \times b \times c) \rightarrow d) \times c) \rightarrow ((a \times b) \rightarrow d)$ & = & $\lambda(f,z). \lambda(x,y). f(x,y,z)$\\

& & & $(((a \times b \times c) \rightarrow d) \times a \times b) \rightarrow (c \rightarrow d)$ & = & $\lambda(f,x,y). \lambda(z). f(x,y,z)$\\
& & & $(((a \times b \times c) \rightarrow d) \times b \times c) \rightarrow (a \rightarrow d)$ & = & $\lambda(f,y,z). \lambda(x). f(x,y,z)$\\
& & & $(((a \times b \times c) \rightarrow d) \times a \times c) \rightarrow (b \rightarrow d)$ & = & $\lambda(f,x,z). \lambda(y). f(x,y,z)$\\
& & & $(((a \times b \times c) \rightarrow d) \times a \times b \times c) \rightarrow (\{\} \rightarrow d)$ & = & $\lambda(f,x,y,z). \lambda(). f(x,y,z)$\\
\noalign{\smallskip}\hline
\end{tabular}}
\label{tab:signatures}
\end{table}

Additional concepts from programming languages in this model are 
partial function application
({\scriptsize $PFA_1$}),
partial function application $n$ ({\scriptsize $PFA_n$}),
single-argument partial
evaluation ({\scriptsize $SPE$}), and partial evaluation
({\scriptsize $PE$})~\cite{introPartialEvaluation}.
Partial function application, \texttt{papply1}, takes a function and its first
argument and returns a function accepting the remainder of its parameters.  The
function \texttt{papplyn}, on the other hand, takes a function $f$ and all of
the first $n$ of $m$ arguments to $f$ where $n \leqslant m$, and returns a
function accepting the remainder of its $(m-n)$ parameters.  Notice that with
single-argument partial evaluation, the input function may be partially
evaluated with only one argument at a time.  These concepts correspond to
higher-order functions that each take a function and some subset of its
parameters as arguments.  The type signatures for these functions are given in
Table~\ref{tab:signatures}.  All of these functions except \texttt{apply}
return a function.  Table~\ref{tab:definitions} provides definitions of
\texttt{papply1}, \texttt{papplyn}, and \texttt{smix} in Scheme.
We use the symbol \texttt{mix} from~\cite{introPartialEvaluation} to denote the
partial evaluation operation because partial evaluation involves a
\texttt{mix}ture of interpretation and code generation.
The \texttt{mix} operator accepts two arguments: a function to be partially evaluated
and a static assignment of values to any subset of its parameters.  The
semantics of the expression $[\![\mathtt{f}]\!] \mathtt{3}$ in the notation
from~\cite{introPartialEvaluation} are `invoke \texttt{f} on \texttt{3}' or
\texttt{f(3)}.  Consider a function \texttt{pow} that accepts a \texttt{base}
and an \texttt{exponent}, in that order, as arguments and returns the
\texttt{base} raised to the \texttt{exponent}.  The semantics of the expression
$\mix [\mathtt{pow}, \mathtt{exponent=2}]$ are `partially evaluate \texttt{pow}
with respect to \texttt{exponent} equal to two,' an operation which returns
$\mathtt{pow}_{\mathtt{exponent=2}}$ that accepts only a \texttt{base} (i.e., a
squaring function).  Therefore, {\scriptsize $[\![ \overbrace{\mix
[\mathtt{pow},
\mathtt{exponent=2}]}^{\mathrm{a~partial~evaluation,}~\mathtt{pow}_{\mathtt{exponent=2}}}]\!]
\mathtt{3} = \overbrace{[\![ \mathtt{pow} ]\!] [\mathtt{3},
\mathtt{2}]}^{\mathrm{a~complete~evaluation}} = 9$}. Given a ternary function
\texttt{f} with integer parameters \texttt{x}, \texttt{y}, and \texttt{z}:
{\scriptsize
$\mathtt{f}_{\mathtt{y=2}} = \mix [\mathtt{f}, \mathtt{y=2}]$}
and {\scriptsize $[\![\mathtt{f}]\!] [\mathtt{1}, \mathtt{2}, \mathtt{3}] =
[\![\mix [\mathtt{f}, \mathtt{y=2}]]\!] [\mathtt{1}, \mathtt{3}]$}.  In
general, {\scriptsize $[\![\mix [\mathit{f},
\mathit{input}_\mathit{static}]]\!] \mathit{input}_\mathit{dynamic} =
[\![\mathit{f}]\!] [\mathit{input_\mathit{static}},
\mathit{input_\mathit{dynamic}}]$}.

These functions are general in that they accept a function of any arity as
input.  The functions \texttt{curry}, \texttt{papply1}, \texttt{papplyn},
\texttt{smix}, and \texttt{mix} are \textit{closed} operators over their input
set (i.e., they take a function as input and return a function as output).
Here, we are interested in a progressive series of applications of each of
these functions that terminates at a fixpoint.  Therefore, we superscript a
concept mnemonic $X$ in the numerator with a $\star$, where applicable, to
indicate a progressive series of applications of the corresponding function
ending in a fixpoint.  For instance, repeatedly applying \texttt{papplyn} to a
ternary function \texttt{f} as \texttt{(apply (papplyn (papplyn f small) mild)
no)} realizes the episode $\prec$\textrm{size blend cream}$\succ$ in addition
to the $\prec$\textrm{size (blend cream)}$\succ$, $\prec$\textrm{(size blend)
cream}$\succ$, and $\prec$\textrm{(size blend cream)}$\succ$ episodes which are
realized with only a single application of \texttt{papplyn}.
Thus, the expression {\scriptsize
$\frac{PE^{\star}}{\textrm{size blend cream}}$} denotes the set of all six
permutations of \{size, blend, cream\} and all permutations of all set
partitions of \{size, blend, cream\} or, in other words, all thirteen, possible
episodes to
complete the dialog: {\scriptsize \{$\prec$(size blend cream)$\succ$,
$\prec$(size blend) cream$\succ$, $\prec$cream (size blend)$\succ$, 
$\prec$(blend cream) size$\succ$, $\prec$size (blend cream)$\succ$,
$\prec$(size cream) blend$\succ$, $\prec$blend (size cream)$\succ$,
$\prec$size blend cream$\succ$, \dots remaining five
permutations of \{size, blend, cream\} \dots \}}.

This notation also contains a prime ($'$) superscript.  While the star
($\star$) superscript permits repeated applications, but does not require them,
the prime ($'$) superscript requires repeated applications of the operator
until a fixpoint is reached. For instance, the episode 
{\scriptsize \{$\prec$\textrm{size
(blend cream)}$\succ$} is specified by {\scriptsize $\frac{{SPE}^{\star}}{\textrm{size blend
cream}}$}, but not by {\scriptsize $\frac{{SPE}^{'}}{\textrm{size blend cream}}$}.

Note that {\scriptsize $\frac{C}{\textrm{size blend cream}} =
\frac{{C}^{\star}}{\textrm{size blend cream}}$} because the function returned
from the application of a curried function is already in curried form; there is
no reed to recurry it.  In general, {\scriptsize $\frac{C}{\textrm{\dots}} =
\frac{{C}^{\star}}{\textrm{\dots}}$} for any denominator \dots\ common to both
expressions. When the denominator is irrelevant to the discussion at hand we
drop it and simply use only the concept mnemonic to refer to a set of episodes.
Thus, 
{\scriptsize 
$C = C^{\star}$}.  Also,
{\scriptsize
$I=I^*$} since \texttt{apply} does not return a function.
However, we can superscript
{\scriptsize
$PFA_1$, $PFA_n$, $SPE$}, and
{\scriptsize
$PE$} with a
{\scriptsize
$\star$}
symbol.

\begin{table}
\centering
\caption{Definitions of (left) \texttt{papply1},
(center) \texttt{papplyn}, and (right) \texttt{smix} in Scheme.}
{\scriptsize
\renewcommand{\tabcolsep}{0cm}
\begin{tabular}{lll}
\hline\noalign{\smallskip}

\begin{tabular}{l}
\begin{lstlisting}[stepnumber=0]
(define papply1
  (lambda (fun arg)
    (lambda x
      (apply fun (cons arg x)))))
\end{lstlisting}
\end{tabular} &

\begin{tabular}{l}
\begin{lstlisting}[stepnumber=0]
(define papplyn
  (lambda (fun . args)
    (lambda x
      (apply fun (append args x)))))
\end{lstlisting}
\end{tabular} &

\begin{tabular}{l}
\begin{lstlisting}[stepnumber=0]
(define smix
   (lambda (fun static_arg)
      (mix fun static_arg)))
\end{lstlisting}
\end{tabular}\\

\noalign{\smallskip}\hline
\end{tabular}}
\label{tab:definitions}
\end{table}

The second row of Table~\ref{tab:posets} and the third row
of Table~\ref{tab:others}
shows enumerated specifications of
dialogs for ordering coffee, among others.
Each dialog is also specified using this notation
based on these concepts from programming languages in the 
fourth row of Table~\ref{tab:posets} and
the second row of Table~\ref{tab:others}.
We associate a fixed dialog with \textit{currying}
({\scriptsize$C$}) (fourth row of column (a) in Table~\ref{tab:posets})
and a complete, \mi\ dialog with \textit{partial evaluation}
({\scriptsize $PE^{\star}$})
(fourth row of column (e) in Table~\ref{tab:posets}).
The concepts from programming
languages in the second row in Table~\ref{tab:others}
(and combinations of them) within the
context of an expression in this notation help specify dialogs between the
fixed and complete, \mi\ ends of this dialog spectrum,
shown in Table~\ref{tab:posets}.  Note that the order of
the terms in the denominator matters (i.e., {\scriptsize $(\frac{C}{\textrm{a b
c}} = \{\prec\textrm{a b c}\succ\}) \neq (\frac{C}{\textrm{b a c}} = \{\prec
\textrm{b a c}\succ\})$)}.
Also, note that when the number of questions posed in a dialog is less than
three, expressions with a
{\scriptsize
$C$} or
{\scriptsize 
$PFA_1$}
in the numerator specify the same
episode (e.g., {\scriptsize $\frac{C}{\textrm{a b}} = \frac{PFA_1}{\textrm{a
b}} =$ \{$\prec$a b$\succ$\})}.
The concept mnemonic in each cell of
Table~\ref{tab:pracspace} connect the dialogs in that cells
with the concept from programming languages used to
specify them.

Note that there is always only one episode possible
in any dialog specified using only
one of the 
{\scriptsize
$I$, $C$}, or
{\scriptsize
$PFA_1$}
mnemonics in the numerator.
There are always $q$ episodes
in any dialog specified using only one of the
{\scriptsize
$PFA_n$} or
{\scriptsize
$SPE$}
mnemonics in the numerator,
where $q$ is the
number of questions posed in a dialog.  The number of episodes in any dialog
specified using any of the
{\scriptsize
$PFA_{n}^{\star}$, $SPE^{'}$, $PE$,}
or
{\scriptsize
$PE^{\star}$} mnemonics in the numerator as a function of $q$ is
{\scriptsize $2^{q-1}$, $q!$, $\sum_{p=1}^{q}\binom{q}{p}$}, and
{\scriptsize $\sum_{p=1}^{q}p!\times S(q,~p)$}\footnote{Let {\scriptsize $s(m, n)$} be
the set of all partitions of a set of size $m$ into exactly $n$ non-empty
subsets, where $n$ is a positive integer and $n \leqslant m$.
The \textit{Stirling number} of a set of size $m$ is {\scriptsize $S(m, n) = |s(m,
n)|$}.}, respectively.
The episodes in any dialog specified using only one of the
{\scriptsize
$I$, $C$, $PFA_1$,
$PFA_n$,
$PFA_{n}^{\star}$,
$SPE$, $SPE^{'}$, 
$PE$}, or
{\scriptsize
$PE^{\star}$} mnemonics in the numerator are
related to each other in multiple ways.  For instance, by definition of the
{\scriptsize
$\star$} symbol here,
{\scriptsize
$X \subseteq X^{\star}$}, where
{\scriptsize
$X$} is any mnemonic corresponding to 
a concept from programming languages used in this model (e.g.,
{\scriptsize
$PFA_1$} or
{\scriptsize
$PE$}). Other relationships include:
{\scriptsize
$PFA_n~\cup~SPE \subset PE$,
$((PFA_{n}^{\star}-PFA_n)~\cup~(SPE^{\star}-SPE))~\cap~PE~=~\emptyset$,
$PFA_1 \subset PFA_n$,
$I \cup C \cup PFA_1 \cup PFA_n$ 
$\subset PFA_{n}^{\star}$, and
$PFA_{n}^{\star} \subset PE^{\star}$}.
Lastly,
{\scriptsize
$I~\cup~C~\cup~PFA_1~\cup~PFA_n~\cup~PFA_{n}^{\star}~\cup~SPE~\cup~SPE^{'}~\cup~PE~\subset~PE^{\star}$}
meaning that partial evaluation subsumes other all other concepts in this
model.  The implication of this, as we see in the following section, is that
any dialog specified using this notation can be supported through partial
evaluation.

This notation is expressive enough to also capture dialogs involving multiple
sub-dialogs, by nesting these expressions in the denominator. For 
instance, consider
{\scriptsize $\frac{SPE^{'}}{\frac{{PE}^{\star}}{\textrm{cream sugar}}
\frac{{PE}^{\star}}{\textrm{eggs toast}}}$},
where the user can specify coffee and
breakfast choices in any order, and can specify the sub-parts of coffee and
breakfast in any order, but cannot mix the atomic responses of the two
(e.g., the $\prec$cream eggs sugar toast$\succ$ episode is not permitted).

\subsection{Evaluation}

We denote the space of dialogs possible given $q$, the number of questions
posed in a dialog, with the symbol
{\scriptsize
$\mathcal{U}_{q}$}.
Let
{\scriptsize
$X$} denote a concept from programming languages in this model (e.g.,
{\scriptsize
$C$} or
{\scriptsize
$PE^{\star}$}).  We use the symbol
{\scriptsize
$\mathcal{X}$} to denote a \textit{class} of
dialogs (e.g.,
{\scriptsize
$\mathcal{C}$} or
{\scriptsize
$\mathcal{PE}^{\star}$}), where a class is
a set of dialogs where each dialog in the set can be
specified with concept mnemonic
{\scriptsize
$X$} in the numerator
and $q$ is the number of questions posed in each dialog.
The number of dialogs possible given a value for $q$ is 
{\scriptsize
$|\mathcal{U}_{q}| = 2^{|PE^{\star}|}-1$}.
Of all of those dialogs,
there are
{\scriptsize $2^{|PE^{\star}|}-3q!-q-5$} dialogs\footnote{For $q \geqslant 3$,
{\tiny
$|\Delta| = |\mathcal{U}| - |\mathcal{I}| - |\mathcal{C}| -
|\mathcal{PFA}_1| -
|\mathcal{PFA}_n| - |\mathcal{PFA}_{n}^{\star}| - |\mathcal{SPE}| -
|\mathcal{PE}| -
|\mathcal{PE}^{\star}|$}.
The cases where $q=1$ and $q=2$ are the only
cases where
{\tiny
$|\Delta| \ne |\mathcal{U}| - |\mathcal{I}| - |\mathcal{C}| -
|\mathcal{PFA}_1| -
|\mathcal{PFA}_n| - |\mathcal{PFA}_{n}^{\star}| - |\mathcal{SPE}| -
|\mathcal{PE}| -
|\mathcal{PE}^{\star}|$}.
This is because when $q=1$, the one
specification in each of the individual classes is the same in each class.
Similarly, when $q=2$ some specifications are multi-classified.}
that cannot be specified with a single 
concept (e.g.,
dialogs b, c, and d in Table~\ref{tab:posets}).
We refer to the class containing these dialogs as
{\scriptsize
$\Delta$}.
There are notable observations on any space
{\scriptsize
$\mathcal{U}_{q}$}:
\begin{inparaenum}[\itshape a\upshape)] \item its
classes are totally disjoint, \item the
{\scriptsize
$\mathcal{I}$, $\mathcal{SPE}$, $\mathcal{SPE}^{'}$,
$\mathcal{PE}$}, and
{\scriptsize
$\mathcal{PE}^{\star}$} classes always
contain only one specification given any $q$,
\item the
{\scriptsize
$\mathcal{PFA}_1$}
class always contains $q$ specifications because there exists one specification
per question, where the response to that question is supplied first and the
responses to all remaining questions arrive next in one utterance,
\item
{\scriptsize
the $\mathcal{C}$,
$\mathcal{PFA}_n$}, and
{\scriptsize
$\mathcal{PFA}_{n}^{\star}$} classes always contain $q!$ specifications as each
contains one specification for each episode in a dialog specified
using the 
{\scriptsize
$SPE^{'}$} concept, and
\item therefore, the number of dialogs specifiable with
only a single instance of a concept mnemonic is
{\scriptsize
$3q! + q + 5$}.
\end{inparaenum}

However, this notation for dialog specification is
expressive enough to specify the dialogs in the 
{\scriptsize
$\Delta$} class
because those dialogs can be expressed as either a union of the
set of episodes specified by more than one dialog
specification called a \textit{compound expression} (e.g., dialog
c in Table~\ref{tab:posets}, or {\scriptsize $\frac{I}{\textrm{x y z}} \cup
\frac{PFA_1}{\textrm{x y z}} =$ \{$\prec$(x y z)$\succ$, $\prec$x (y
z)$\succ$\})} or expressed as dialogs involving
\textit{sub-dialogs} through the use of
nesting~\cite{vt-staging} (e.g., dialogs b and d in Table~\ref{tab:posets}),
or both (e.g., {\scriptsize
$\frac{C}{\textrm{size} \frac{SPE}{\textrm{blend cream}}} \cup
\frac{C}{\textrm{blend} \frac{SPE}{\textrm{cream size}}} \cup
\frac{C}{\textrm{cream blend size}}$}).

In dialogs containing two or more terms
in the denominator, where at least one of the terms is a sub-dialog
(e.g., dialogs {\scriptsize $\frac{...}{\textrm{a} \frac{C}{\textrm{b c}}}$} and
{\scriptsize $\frac{...}{\frac{C}{\textrm{a b}} \frac{{PE}^{\star}}{\textrm{c d}}}$}, but not
{\scriptsize $\frac{...}{\frac{I}{\textrm{a b}}}$}), each of the
{\scriptsize
$I$, $PFA_n$, $PFA_{n}^{\star}$, $PE$}, and
{\scriptsize
$PE^{\star}$} concept mnemonics is
not a candidate for the numerator.  This is
because those concepts require (in the case of
{\scriptsize
$I$}) or support multiple
responses per utterance and it is not possible to complete multiple
sub-dialogs in a single utterance or complete
a sub-dialog and an individual question in a single utterance.
The 
{\scriptsize
$PFA_1$} and
{\scriptsize
$SPE$} concept mnemonics only suffice for two categories
of dialogs containing sub-dialogs: those with no more than
two terms in the denominator, where one of the terms is a sub-dialog
(e.g.,
{\scriptsize $\frac{PFA_1}{\textrm{a} \frac{{PE}^{\star}}{\textrm{b c}}}$},
{\scriptsize $\frac{PFA_1}{\frac{{PE}^{\star}}{\textrm{a b}} \frac{{PE}^{\star}}{\textrm{c
d}}}$},
{\scriptsize $\frac{SPE}{\textrm{a} \frac{{PE}^{\star}}{\textrm{b c}}}$, and
$\frac{SPE}{\frac{{PE}^{\star}}{\textrm{a b}} \frac{{PE}^{\star}}{\textrm{c
d}}}$})
and those with more than two terms in the
denominator where only the first term is a sub-dialog (e.g.,
{\scriptsize $\frac{PFA_1}{\frac{{PE}^{\star}}{\textrm{a b}} \textrm{c d e f}}$ and
$\frac{SPE}{\frac{{PE}^{\star}}{\textrm{a b}} \textrm{c d e f}}$}).
This is because when used as the numerator in an expression whose denominator
contains more than two terms, one of which is a sub-dialog not in the
first position,
{\scriptsize $PFA_1$} and
{\scriptsize $SPE$} require multiple responses in
the second and final utterance.
Hence,
{\scriptsize
$C$} is the only concept mnemonic that can always be used in the numerator of an
expression containing any number of sub-dialogs in the denominator.
However, 
{\scriptsize
$C$} only supports fixed orders of responses.
Thus, we need a mnemonic for a concept that restricts utterances to one
response and only permits one sub-dialog to be pursed at a time,
but also permits all possible completion orders.  Such a concept could
be used to specify a dialog with more than two terms in the denominator, any of
which can be a sub-dialog, that can be completed in any order.  The concept
represented by the mnemonic
{\scriptsize
$SPE^{'}$} is ideal for this purpose (see column (i) in
Table~\ref{tab:others}).
Note that {\scriptsize $\frac{C}{
\frac{{PE}^{\star}}{\textrm{a b}}
\frac{{PE}^{\star}}{\textrm{c d}}
\frac{{PE}^{\star}}{\textrm{e f}}} \neq
\frac{SPE^{'}}{
\frac{{PE}^{\star}}{\textrm{a b}}
\frac{{PE}^{\star}}{\textrm{c d}}
\frac{{PE}^{\star}}{\textrm{e f}}}$}, and
{\scriptsize
$SPE^{'} \subset PE^{\star}$}.
The form of these expressions in this notation based on concepts from
programming languages can be described by the \text{datatype}
definitions in \textsc{ml} in Listing~\ref{lst:datatypes}.

\begin{lstlisting}[float,frame=tblr,language=ML,stepnumber=0,
caption={Datatype definitions in \textsc{ml}
describing the form of the expressions in the notation based on concepts from
programming languages used in this article.},label=lst:datatypes]
datatype Concept_mnemonic = I | C | PFA1 | PFAN | PFANstar |
                            SPE | SPEprime | PE | PEstar;

(* Concept_mnemonic is numerator and Dialog_spec is denominator *)
datatype Dialog_spec = A_Dialog_spec of Concept_mnemonic * Dialog_spec

(* A dialog specification is a either a single expression or
   a union of single expressions, called a compound expression *)
datatype Dialog_specification =
   A_single_dialog_spec of Dialog_spec | A_union_of_dialog_specs of Dialog_spec list;
\end{lstlisting}

Specifying dialogs in the spectrum shown in Table~\ref{tab:posets} with this
notation based on concepts from programming languages has multiple effects:
\begin{inparaenum}[\itshape a\upshape)] \item it helps bring structure to the
space between the two ends of the spectrum, \item it helps us losslessly
compress the episodes in an enumerated specification of a dialog without
enumerating all of the episodes to capture the possible orders and combinations
of responses therein and, therefore, provides a shorthand notation for dialog
specification, akin to the Hasse diagram method,
and \item a dialog specified in this notation provides a design for
implementing the dialog, as we see below.  \end{inparaenum}

We currently have nine language concepts in our model.
Each concept enriches the expressivity of the notation at capturing dialog
specifications from which to model dialogs.
An attractive consequence of this notation for dialog specification is that the
(nested) structure of the expression, and the language concepts used therein,
provides a blueprint for staging (i.e., implementing) the dialog.  In other
words, the concepts from programming languages are not just helpful metaphors
for dialog specification (as we harness the philosophical analogs between
programming languages and natural languages), but also lend insight into
operationalizing the dialog.

\section{Staging \Mi\ Dialogs by Partial Evaluation}
\label{sec:staging}

\begin{figure}
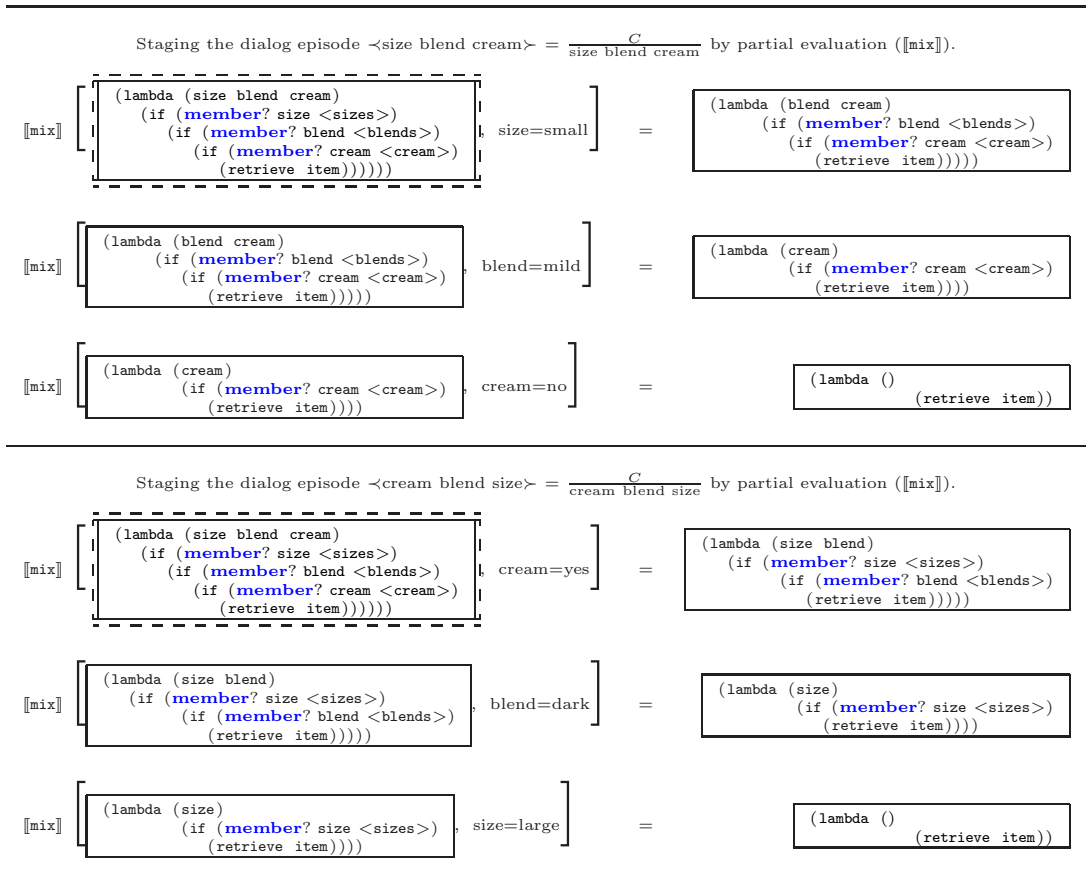

\centering
{\tiny
\begin{tabular}{llr}
\hline\noalign{\smallskip}
\\

\multicolumn{3}{c}{Staging the dialog episode
$\prec$\textrm{size blend cream}$\succ$ $=$
$\frac{C}{\textrm{size blend cream}}$
by partial evaluation ($\mix$).}\\

& &\\

$\mix$ {\Huge [} 
\begin{tabular}{:|l|:}
\hdashline
\hline
{\color{black}
\begin{lstlisting}[stepnumber=0,basicstyle=\tiny]
(lambda (size blend cream)
   (if (member? size <sizes>)
      (if (member? blend <blends>)
         (if (member? cream <cream>)
            (retrieve item))))))
\end{lstlisting}}\\
\hline
\hdashline
\end{tabular},~~size=small{\Huge ]} & = &

\begin{tabular}{|l|}
\hline
\begin{lstlisting}[stepnumber=0,basicstyle=\tiny]
(lambda (blend cream)
      (if (member? blend <blends>)
         (if (member? cream <cream>)
            (retrieve item)))))
\end{lstlisting}\\
\hline
\end{tabular}\\

\medskip
& &\\

$\mix$ 
{\Huge [}\begin{tabular}{|l|}
\hline
\begin{lstlisting}[stepnumber=0,basicstyle=\tiny]
(lambda (blend cream)
      (if (member? blend <blends>)
         (if (member? cream <cream>)
            (retrieve item)))))
\end{lstlisting}\\
\hline
\end{tabular},~~blend=mild{\Huge ]} & = &

\begin{tabular}{|l|}
\hline
\begin{lstlisting}[stepnumber=0,basicstyle=\tiny]
(lambda (cream)
         (if (member? cream <cream>)
            (retrieve item))))
\end{lstlisting}\\
\hline
\end{tabular}\\

\medskip
& &\\

$\mix$ 
{\Huge [}\begin{tabular}{|l|}
\hline
\begin{lstlisting}[stepnumber=0,basicstyle=\tiny]
(lambda (cream)
         (if (member? cream <cream>)
            (retrieve item))))
\end{lstlisting}\\
\hline
\end{tabular},~~cream=no{\Huge ]} & = &

\begin{tabular}{|l|}
\hline
{\color{black}
\begin{lstlisting}[stepnumber=0,basicstyle=\tiny]
(lambda ()
            (retrieve item))
\end{lstlisting}}\\
\hline
\end{tabular}\\

& &\\
\noalign{\smallskip}\hline\noalign{\smallskip}
& &\\

\multicolumn{3}{c}{Staging the dialog episode
$\prec$\textrm{cream blend size}$\succ$ $=$
$\frac{C}{\textrm{cream blend size}}$
by partial evaluation ($\mix$).}\\

& &\\

$\mix$ 
{\Huge [}
\begin{tabular}{:|l|:}
\hdashline
\hline
{\color{black}
\begin{lstlisting}[stepnumber=0,basicstyle=\tiny]
(lambda (size blend cream)
   (if (member? size <sizes>)
      (if (member? blend <blends>)
         (if (member? cream <cream>)
            (retrieve item))))))
\end{lstlisting}}\\
\hline
\hdashline
\end{tabular},~~cream=yes{\Huge ]} & = &
\begin{tabular}{|l|}
\hline
\begin{lstlisting}[stepnumber=0,basicstyle=\tiny]
(lambda (size blend)
   (if (member? size <sizes>)
         (if (member? blend <blends>)
            (retrieve item)))))
\end{lstlisting}\\
\hline
\end{tabular}\\

\medskip
& &\\

$\mix$ 
{\Huge [}\begin{tabular}{|l|}
\hline
\begin{lstlisting}[stepnumber=0,basicstyle=\tiny]
(lambda (size blend)
   (if (member? size <sizes>)
         (if (member? blend <blends>)
            (retrieve item)))))
\end{lstlisting}\\
\hline
\end{tabular},~~blend=dark{\Huge ]} & = &
\begin{tabular}{|l|}
\hline
\begin{lstlisting}[stepnumber=0,basicstyle=\tiny]
(lambda (size)
         (if (member? size <sizes>)
            (retrieve item))))
\end{lstlisting}\\
\hline
\end{tabular}\\

\medskip
& &\\

$\mix$ 
{\Huge [}\begin{tabular}{|l|}
\hline
\begin{lstlisting}[stepnumber=0,basicstyle=\tiny]
(lambda (size)
         (if (member? size <sizes>)
            (retrieve item))))
\end{lstlisting}\\
\hline
\end{tabular},~~size=large{\Huge ]} & = &
\begin{tabular}{|l|}
\hline
{\color{black}
\begin{lstlisting}[stepnumber=0,basicstyle=\tiny]
(lambda ()
            (retrieve item))
\end{lstlisting}}\\
\hline
\end{tabular}\\
\\
\hline
\end{tabular}}
\caption{Staging dialog episodes by partial evaluation,
explicitly illustrating the intermediate output of each partial
evaluation.  Dotted boxes reinforce that
both series of transformations, top half and bottom half, start
with the same script.}
\label{fig:transexample}
\end{figure}

Our notation for specifying \mi\ dialogs lends itself to two methods of dialog
implementation: \begin{inparaenum}[i)] \item using partial evaluation to stage
the interaction~\cite{introPartialEvaluation}, or \item using a set of rewrite
rules to stage the interaction~\cite{termrewriting}\end{inparaenum}.
We use an example to illustrate how dialogs can be staged with partial
evaluation.  Consider the ternary Scheme function shown 
within a dotted border in
Fig.~\ref{fig:transexample},
simplified for purposes of
succinct exposition.\footnote{An expression of the form \texttt{<...>}
is used to represent a list of valid choices (e.g., \texttt{<sizes>}
could represent the list \texttt{`(small medium large)}). Moreover,
these functions being partially evaluated in
Figs.~\ref{fig:transexample} and~\ref{fig:transexample2} omit \texttt{else}
(exceptional) branches
for purposes of succinct exposition.}
Notice that it only
models one dialog episode: $\prec$\textrm{size blend cream}$\succ$.  We define
this function without the intent of ever invoking it, and rather only with the
intent of progressively transforming it automatically with partial evaluation
for the sake of staging the interaction of a \mi\ dialog.  Thus, we only use
this function as a malleable data object, and when it has been completely
consumed through transformation, the dialog is complete.  

The top half of Fig.~\ref{fig:transexample} demonstrates how the
$\prec$\textrm{size blend cream}$\succ$ episode is staged by this process.
The \textit{same} function can be
used to realize a completely different episode than the one after which it is
modeled.  For instance, the bottom half of
Fig.~\ref{fig:transexample}
demonstrates how the $\prec$\textrm{cream blend size}$\succ$ episode can
be staged by this process, with the \textit{same} function.
While Fig.~\ref{fig:transexample} shows how the function is transformed after
each progressive partial evaluation in the process,
Fig.~\ref{fig:transexample2} omits these intermediary outputs and, thus,
provides an alternate view of Fig.~\ref{fig:transexample}.

\begin{figure}
\centering
{\tiny
\begin{tabular}{c}
\hline\noalign{\smallskip}
\\

Staging the dialog episode
$\prec$\textrm{size blend cream}$\succ$
$=$ $\frac{C}{\textrm{size blend cream}}$
by partial evaluation ($\mix$).\\

\\

$\mix$ {\Huge [}

$\mix$ {\Huge [}

$\mix$ {\Huge [} 
\begin{tabular}{:|l|:}
\hdashline
\hline
{\color{black}
\begin{lstlisting}[stepnumber=0,basicstyle=\tiny]
(lambda (size blend cream)
   (if (member? size <sizes>)
      (if (member? blend <blends>)
         (if (member? cream <cream>)
            (retrieve item))))))
\end{lstlisting}}\\
\hline
\hdashline
\end{tabular},~~size=small{\Huge ]},~~blend=mild{\Huge ]},~~cream=no{\Huge ]}

= 

\begin{tabular}{|l|}
\hline
{\color{black}
\begin{lstlisting}[stepnumber=0,basicstyle=\tiny]
(lambda ()
   (retrieve item))
\end{lstlisting}}\\
\hline
\end{tabular}\\

\hline

Staging the dialog episode
$\prec$\textrm{cream blend size}$\succ$
$=$ $\frac{C}{\textrm{cream blend size}}$
by partial evaluation ($\mix$).\\

$\mix$ {\Huge [}

$\mix$ {\Huge [}

$\mix$ {\Huge [} 
\begin{tabular}{:|l|:}
\hdashline
\hline
{\color{black}
\begin{lstlisting}[stepnumber=0,basicstyle=\tiny]
(lambda (size blend cream)
   (if (member? size <sizes>)
      (if (member? blend <blends>)
         (if (member? cream <cream>)
            (retrieve item))))))
\end{lstlisting}}\\
\hline
\hdashline
\end{tabular},~~cream=yes{\Huge ]},~~blend=dark{\Huge ]},~~size=large{\Huge ]}

=

\begin{tabular}{|l|}
\hline
{\color{black}
\begin{lstlisting}[stepnumber=0,basicstyle=\tiny]
(lambda ()
   (retrieve item))
\end{lstlisting}}\\
\hline
\end{tabular}\\
\\
\hline
\end{tabular}}
\begin{minipage}{\linewidth}
\caption{Alternate view of Fig.~\ref{fig:transexample}, without explicit
illustration of the intermediate outputs of each successive partial
evaluation.\protect\footnote{\tiny It
is not important that the individual responses communicated in these two staged
dialog episodes are different (i.e., \texttt{(small mild no)} and \texttt{(yes
dark large)}).  Rather it is notable that the responses to the questions in
each episode are provided in different orders.  The orders could still be
different even if the responses were the same (e.g., \texttt{(small mild no)}
versus \texttt{(no mild small)}).}}
\label{fig:transexample2}
\end{minipage}
\end{figure}

While the control flow models only one episode (in this case,
{\scriptsize $\prec$\textrm{size blend cream}$\succ$}), through partial evaluating we can
stage the interaction required by thirteen distinct episodes. In general, by
partially evaluating a function representing only one episode, we can realize
{\scriptsize $\sum_{p=1}^{q}p!\times S(q,~p)$} distinct episodes
(i.e., {\scriptsize $|\mathcal{D}_{cmi}|$}), where $q$ is
the number of questions posed in a dialog, and {\scriptsize $S(m,n)$} is size
of the set of all partitions of a set of size $m$ into exactly $n$ non-empty
subsets, where $n$ is a positive integer and $n \leqslant m$ (i.e., the
\textit{Stirling number} of a set of size $m$).
This `model one episode, stage multiple' feature is a significant result of our
approach to dialog modeling and implementation.

Notice from Fig.~\ref{fig:transexample} that, at any point in the interaction,
a script always explicitly models the questions that remain unanswered and,
therefore, implicitly models the questions that have been answered.  As a
result, it is always clear what information to prompt for next.  In the \mi\
dialog community, keeping track of what has and has not been communicated is
called \textit{dialog management}.

The dialog {\scriptsize $\frac{I}{\textrm{size blend cream}} =$ \{$\prec$(size
blend cream)$\succ$\}} can be staged with partial evaluation as {\scriptsize
$\mix [\mathrm{f}, \mathrm{size=\dots, blend=\dots, cream=\dots}]$}.
Similarly, {\scriptsize $\frac{PFA_1}{\textrm{size blend cream}} =$
\{$\prec$size (cream blend)$\succ$\}} can be staged with partial evaluation as
{\scriptsize $\mix [\mix [\frac{PFA_1}{\textrm{size blend cream}},
\mathrm{size=\dots}],$ $\mathrm{blend=\dots, cream=\dots}]$}.  The fifth (last)
row of Table~\ref{tab:posets} and the fourth (last) row of
Table~\ref{tab:others} details how dialogs specified using only one of each
concept mnemonic in an expression are staged by partial evaluation, which
subsumes all of the other concepts based on the arguments with which you
partially evaluate.  For instance, {\scriptsize $PFA_{n}^{\star}$} is achieved
by progressively partially evaluating with any prefix of its arguments (see
last row, third (g) column in Table~\ref{tab:others}). 

Given a specification of a dialog in our notation, an alternate implementation
approach involves the use of rewrite rules to stage the
interaction~\cite{termrewriting}. 
The concepts {\scriptsize $I$} and {\scriptsize $C$} are primitive in that any
dialog modelable with our notation
can be represented using \textit{only} {\scriptsize $I$} or {\scriptsize $C$}
concept mnemonics in a specification expression.
In particular, to specify any dialog in the spectrum shown in
Table~\ref{tab:posets} using this notation we can simply translate each episode
in its enumerated specification
as a sub-expression with either an {\scriptsize $I$}
or {\scriptsize $C$}
in the numerator and the entire specification as a union of those
sub-expressions.
As a result, all dialogs specified using this notation can be
reduced to a specification using only the
{\scriptsize
$I$} or
{\scriptsize
$C$} concept mnemonics.
For instance, {\scriptsize $\frac{I}{\textrm{x y z}} \cup
\frac{C}{\textrm{x y z}} \cup \frac{C}{\textrm{y z x}} \cup \frac{C}{\textrm{z
x y}} \cup \frac{C}{\textrm{x} \frac{I}{\textrm{y z}}} =$ \{$\prec$(x y
z)$\succ$, $\prec$x y z$\succ$, $\prec$y z x$\succ$, $\prec$z x y$\succ$,
$\prec$x (y z)$\succ$\}}.
Therefore, we define
\textit{rewrite rules}, not shown here, akin to those in~\cite{WWW2004}, and
can progressively apply them after every utterance, rather than partial
evaluation itself, to transform the representation of the dialog, to stage it.
For instance, the above dialog {\scriptsize
$\frac{PFA_1}{\textrm{size blend cream}}$
$=$ \{$\prec$size (blend cream)$\succ$\}} can
be staged with term rewriting as
{\scriptsize
$\frac{PFA_1}{\textrm{size blend cream}} = \frac{C}{\textrm{size}
\frac{I}{\textrm{blend cream}}}$}
(first rewrite), and {\scriptsize
$[\frac{C}{\textrm{size} \frac{I}{\textrm{blend cream}}},
\mathrm{size=\dots}] =
\frac{C}{\frac{I}{\textrm{blend cream}}} = \frac{I}{\textrm{blend cream}}$}
(second rewrite),
and
{\scriptsize $[\frac{I}{\textrm{blend cream}},$
$\mathrm{blend=\dots, cream=\dots}] =~\sim$}
(i.e., dialog complete).  Similarly, {\scriptsize
$[\frac{SPE^{'}}{\frac{PE}{\textrm{a b}}
\frac{PE}{\textrm{c d}}}, \mathtt{d}=\dots] =
\frac{SPE^{'}}{\frac{PE}{\textrm{c}} \frac{PE}{\textrm{a b}}} =
\frac{SPE^{'}}{\frac{C}{\textrm{c}} \frac{PE}{\textrm{a b}}} =
\frac{C}{\textrm{c} \frac{PE}{\textrm{a b}}}$}, and
{\scriptsize $[\frac{C}{\textrm{c}
\frac{PE}{\textrm{a b}}}, \mathtt{c}=.] = \frac{C}{\frac{PE}{\textrm{a b}}}
= \frac{PE}{\textrm{a b}}$}, and
{\scriptsize $[\frac{PE}{\textrm{a b}}, \mathtt{b}=\dots] =
\frac{PE}{\textrm{a}} = \frac{C}{\textrm{a}} = \frac{I}{\textrm{a}}$}, and
finally {\scriptsize $[\frac{I}{\textrm{a}}, \mathtt{a}=\dots] =~\sim$}.

Since partial evaluation can be used to partially apply a function with respect
to \textit{any} subset of its parameters (i.e., it supports the partial
application of a function with all possible orders and combinations of its
arguments), we can stage any unsolicited reporting, \mi\ dialog in this space
using only partial evaluation.  In other words, partial evaluation is a
generalization of any concept from programming languages presented here. A
dialog specification expressed in this notation containing a concept mnemonic
other than {\scriptsize $PE^{\star}$}
represents a particular type of restriction on partial evaluation.

\section{Implementing Dialogs with Partial Evaluation}
\label{sec:implementation}

\begin{figure}
\centering
\includegraphics[scale=0.32]{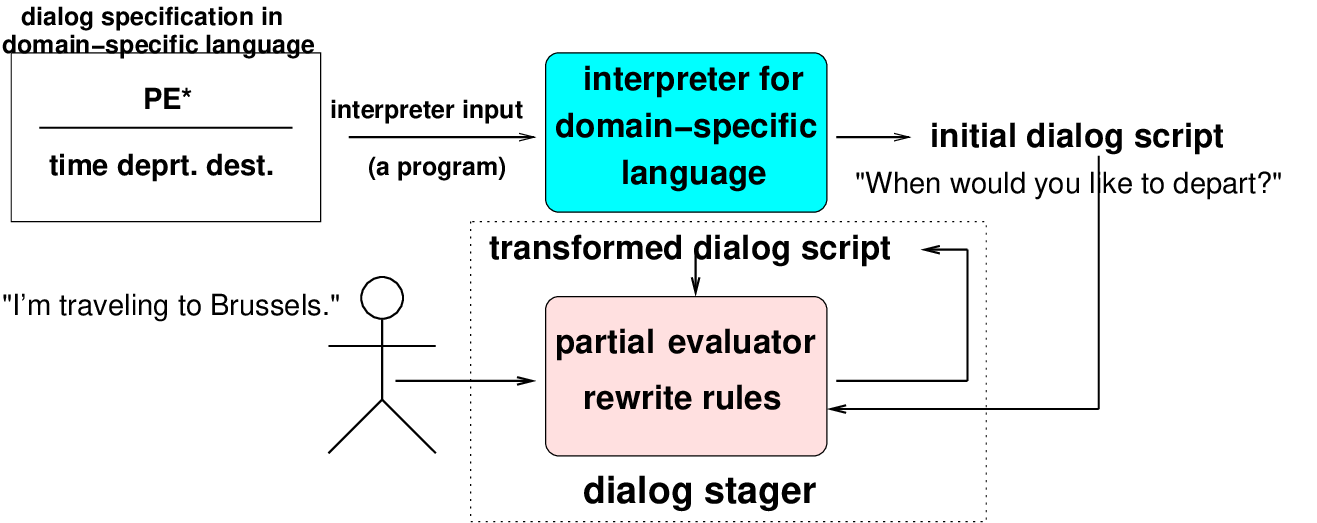}
\caption{Conceptual design of dialog system.}
\label{fig:dsl}
\end{figure}

Guided by this computational model for \mii, we built a dialog modeling and
management toolkit and a dialog engine grounded in these principles.  A
specification of a dialog in this notation, based on concepts from programming
languages, provides a plan for the implementation of the dialog. In this
section we discuss the implementation details of automatically generating a
stager from an enumerated specification of a dialog to be implemented (see
Fig.~\ref{fig:overview}).

\subsection{\Mi\ Dialog Toolkit and Engine}

Our toolkit automates the construction of a \textit{stager} (i.e., a system
implementing a \mi\ dialog) from a specification of a \mi\ dialog in our
notation.  The toolkit can be thought of as an interpreter for a
domain-specific language (i.e., our dialog notation); see 
Fig.~\ref{fig:dsl}.  The prototype, proof-of-concept implementation of both the
toolkit and engine was done in Scheme (described here).

\subsection{Dialog Mining}

Extracting a
minimal specification in this notation, based on concepts from programming
languages, from an enumerated dialog specification is a process we call
\textit{dialog mining}; see transition from the left to the center of
Fig.~\ref{fig:overview}.  While the details of dialog mining are beyond the scope
of this paper, and more appropriate for a data mining audience, we make some
cursory remarks.

\begin{lstlisting}[float,basicstyle=\tiny,frame=tblr,caption={Transcript of an
interactive session with the dialog miner illustrating
how compressed specifications in this notation,
based on concepts from programming
languages, including specifications
a, b, and e in Table~\ref{tab:posets},
are mined from enumerated specifications.},label=lst:dialogmining1]
;; only one utterance; interpretation or complete evaluation
> (mine-expr '(((credit-card grade receipt))))
(("I" credit-card grade receipt))

;; example a: a fixed dialog specification;
;; totally-ordered with only a single response per utterance;
;; currying
> (mine-expr '((credit-card grade receipt)))
(("C" credit-card grade receipt))

;; totally ordered with multiple responses per utterance; partial function application n*
> (mine-expr '((size blend cream) ((size blend) cream)
                 (size (blend cream)) ((size blend cream))))
(("PFA_n*" size blend cream))

;; partially ordered with only a single response per utterance;
;; single-argument partial evaluation'
> (mine-expr (permutations '(size blend cream)))
(("SPE'" size blend cream))

;; example e: a complete, mixed-initiative dialog;
;; partially ordered with multiple responses per utterance;
;; partial evaluation*
> (mine-expr (append (permutations '(size blend cream))
                       '(((size blend) cream)
                         (cream (size blend))
                         (size (blend cream))
                         ((blend cream) size)
                         ((size cream) blend) 
                         (blend (size cream)) 
                         ((size blend cream)))))
(("PE*" blend cream size))

;; example b: a dialog specification containing
;; an embedded, complete, mixed-initiative sub-dialog
> (mine-expr '((PIN account transaction amount) (PIN transaction account amount)))
(("C" PIN ("SPE'" account transaction) amount))
\end{lstlisting}

\begin{lstlisting}[float,basicstyle=\tiny,firstnumber=38,frame=tblr,caption={Continuation of
transcript of an interactive session with the dialog miner
in Listing~\ref{lst:dialogmining1} illustrating
how compressed specifications in this notation, based on concepts from
programming languages,
including specifications c and d in Table~\ref{tab:posets},
are mined from enumerated specifications.},label=lst:dialogmining2]
;; example c: a dialog specification containing an embedded, fixed sub-dialog,
;; or the union of two fixed dialog specifications
> (mine-expr '((receipt sandwich beverage dine-in/takeout)
                 (dine-in/takeout sandwich beverage receipt)))
(("C" receipt sandwich beverage dine-in/takeout)
 ("C" dine-in/takeout sandwich beverage receipt))

;; example d: a dialog specification containing
;; two embedded, complete, mixed-initiative sub-dialogs;
> (mine-expr '((cream sugar eggs toast) (cream sugar toast eggs) (sugar cream eggs toast) 
                (sugar cream toast eggs) (eggs toast sugar cream)(eggs toast cream sugar)
                (eggs toast (cream sugar)) (toast eggs sugar cream) (toast eggs cream sugar)
                (toast eggs (cream sugar)) ((cream sugar) eggs toast)
                ((cream sugar) toast eggs) 
                (cream sugar (eggs toast)) (sugar cream (eggs toast))
                ((cream sugar) (eggs toast))
                ((eggs toast) (cream sugar)) ((eggs toast) cream sugar)
                ((eggs toast) sugar cream)))
(("SPE'" ("PE*" cream sugar) ("PE*" eggs toast)))

;; a dialog specification as the union of three sub-expressions (a compound expression)
> (mine-expr '((size blend cream) (size cream blend) (blend cream size)
                 (cream blend size) (blend size cream)))
(("C" ("SPE'" size blend) cream)
      ("C" size cream blend) ("C" ("SPE'" blend cream) size))

;; case demonstrating the incompleteness of heuristic
;; output should be (``SPE'" x (``C" y z))
> (mine-expr '((x y z) (y z x)))
(("C" x y z) ("C" y z x))
\end{lstlisting}

We designed a recursive, heuristic-based algorithm to address this problem and
implemented it in Scheme.
Listings~\ref{lst:dialogmining1} and \ref{lst:dialogmining2} provide a transcript of an
interactive session with our dialog mining system.  The input to the miner is
an enumerated dialog specification expressed as a list of episodes, where each
episode is also expressed as a list (e.g., see lines 24--31 of
Listing~\ref{lst:dialogmining1}).  The output is a dialog specification in this
notation based on concepts from programming languages.  The form of the output
is a list of lists where each list in the output list represents an expression
in this notation (e.g.,
see line 32 of
Listing~\ref{lst:dialogmining1} and lines 42--43 of Listing~\ref{lst:dialogmining2}).
The \texttt{car} of each list in the output list is the numerator of the
dialog specification in this notation based on concepts from programming
languages and the \texttt{cdr} of it is the denominator.  The nesting of each
list in the output list reflects the nesting in the specification of the
dialog in this notation. For
instance, line 37 of Listing~\ref{lst:dialogmining1} represents
{\scriptsize 
$\frac{C}{\textrm{PIN} \frac{SPE^{'}}{\textrm{account transaction}}
\textrm{amount}}$}.  If the output list contains more than one list (e.g.,
dialog c in Table~\ref{tab:posets} whose compressed specification is shown on
lines 42--43 of Listing~\ref{lst:dialogmining2}), the union of those lists
specifies the dialog.  The miner runs in
\textit{DrRacket}\footnote{\url{http://racket-lang.org/}.} (version 6.0) with
the language set to `Determine language from source'.

Our heuristic is sound in that it always returns a specification in this
notation that represents the input
dialog (i.e., it never returns a wrong answer).  However, it is incomplete in
that it does not always return a minimal specification, where a minimal
specification is one with a minimal number of union operators.  If it cannot
mine a minimal specification, it returns a union of the input set of episodes,
each represented as a sub-expression with
a {\scriptsize $C$} in the numerator.
For instance, line 67 of Listing~\ref{lst:dialogmining2} should display only one
expression (i.e.,
{\scriptsize 
$\frac{SPE^{'}}{\textrm{x}\frac{C}{\textrm{y z}}}$}), but
shows a compound expression containing two sub-expressions instead (i.e.,
{\scriptsize 
$\frac{C}{\textrm{x y z}} \cup \frac{C}{\textrm{y z x}}$}).  We are performing
an evaluation of our heuristic to measure the fraction of
dialog specifications in the universe of possible unsolicited reporting, \mi\
dialogs {\scriptsize $\mathcal{U}$}
for which it is unable to find a minimal specification in
this notation.

We can generalize this problem to one of finding a minimum set of posets
capturing the requirements of a dialog from an enumerated specification of the
dialog.  Formally, we state the problem as:

{\scriptsize
\begin{dialogue}
\speak{Input} A set of posets $P$, all defined over the same set, where the
union of the linear extensions from each poset in $P$ is $L$.
\speak{Output} A minimum set of posets $R$ such that $|R| \leqslant |P|$ and
the union of the linear extensions from each poset of $R$ is $L$.
\end{dialogue}}

\noindent We are currently working on an \textsc{np}-complete proof of this
problem using a reduction to Vertex Cover.  While the dialog mining part of
this work is no less important, here we focus on staging by partial evaluation
and stager generation because it is the aspect of this research most relevant
to the automated software engineering community.

\subsection{Stagers}
\label{subsec:stagers}

\begin{lstlisting}[float,basicstyle=\tiny,frame=tblr,
stepnumber=0,caption={A stager for complete,
\mi\ dialogs (i.e., those
in the {\scriptsize $\mathcal{PE}^{\star}$} class),
simplified for purposes of presentation.},label=lst:PEstarstager]
(define stager_PE*
   (lambda (script)
      (if (not (null? script))
         (let* ((utterance (prompt-for-input))
                (static-input (marshal-utterance-into-a-set-of-parameter/value pairs))
                (specialized-script (mix script static-input)))
            (stager_PE* specialized-script)))))
\end{lstlisting}

Note that the last row of Tables~\ref{tab:posets} and \ref{tab:others}
only demonstrates how to stage dialogs conforming to
each concept from programming languages presented
here. In this section, we address how to implement
stagers with partial evaluation for dialogs that cannot be specified with a
single concept (e.g., dialog b in Table~\ref{tab:posets}) or with a non-compound
expression in this notation (e.g., dialog c in Table~\ref{tab:posets}).
One method of implementing a complete, \mi dialog is to enumerate all possible
ways to complete the dialog as all possible control flows through the
implementation.  For instance, consider dialog e in Table~\ref{tab:posets}, which
is a complete, \mi\ dialog.  Enumerating all possible control flows through that
dialog involves explicitly modeling all of the thirteen separate totally
ordered sets, one for each episode in the specification.
This approach quickly becomes unwieldy even for dialogs with only a few
questions as we demonstrate by capturing the number of episodes in an
enumerated complete, \mi\ dialog specification as a function of the questions
posed therein.
Since an enumerated specification of a complete, \mi\
dialog contains episodes corresponding to all possible permutations of all
possible partitions of the set of questions in the dialog, we define its size,
{\scriptsize $|\mathcal{D}_{cmi}|$}, as the total function,
{\scriptsize $\mathbb{N} \rightarrow
\mathbb{N}$}, equal to {\scriptsize 
$\sum_{p=1}^{q}p!\times S(q,~p)$}, which given $q$, the
number of questions posed in a dialog, computes the total number of episodes
therein.  Therefore, the number of episodes in a complete, \mi\ dialog
specification explodes combinatorially as the number of questions $q$ in the
dialog increase.  For this reason,
we seek to obviate the need to extensionally hardcode
all possible episodes in the control flow of the 
implementation and, thus, improve the control
complexity of dialog implementation, by using partial evaluation to
intensionally support those episodes.

Consider the stager given in Listing~\ref{lst:PEstarstager}.  This
stager, when passed the script shown in
Fig.~\ref{fig:transexample}
does not
need to anticipate when the user is deviating from the only hardwired episode
in the script, by virtue of partial evaluation.  It does not check that
the order of utterances or number of the responses in an utterance conform to
the dialog specification because all orders and combinations are possible.

While complete, \mi\ dialogs can be staged efficiently using this approach,
such dialogs represent only a small fraction of all possible dialogs (i.e.,
there is only one such dialog, given a fixed number of questions).  Most dialog
specifications contain less episodes than those that can be modeled by an
expression in this notation with a 
{\scriptsize $PE^{\star}$} in the numerator. 
Indiscriminately partially evaluating a script such as that shown in
Fig.~\ref{fig:transexample}
to stage dialogs specified by only a {\scriptsize $C$},
as shown in column (a) in Table~\ref{tab:posets},
or {\scriptsize $SPE$},
as shown in column (h) of Table~\ref{tab:others},
realizes \textit{excess} episodes (i.e., episodes
staged that are not in the specification).  On the other hand,
\texttt{apply}ing a script such as that shown in
Fig.~\ref{fig:transexample}
to stage
a dialog conforming to the {\scriptsize $PE^{\star}$}
concept incurs \textit{deficit} (i.e.,
some of the episodes in the specification are not staged).  Using a curried
script to stage a dialog specified by {\scriptsize $SPE$}
yields excess and deficit.  Thus,
while partial evaluation subsumes all other language concepts considered here,
partial evaluation is an `all or nothing' proposition~\cite{MIIMC}. It does not
discriminate against any of the possible partial assignments of input
parameters to the function (i.e., script) being partially evaluated; the script
can be partially evaluated with respect to \textit{any} parameter orders or
combinations.  ``For a dialog script parameterized [by] slot [parameters],
partial evaluation can be used to support all valid possibilities for mixing
initiative, but it cannot restrict the scope of mixing initiative in any way.
In particular this means that, unlike interpretation [or currying], partial
evaluation cannot enforce any individual [episode]''~\cite{MIIMC}.

Thus, to be faithful to a specification, we require a controller to invoke
partial evaluation judiciously with respect to the different orders and
combinations of arguments that reflect the permissible episodes or requirements
of a dialog. We call this controller a \textit{stager} because it stages the
interaction required to complete a dialog. 
Specifically, a stager must restrict the ways in
which partial evaluation is applied to a script in all dialogs except complete,
\mi\ dialogs (i.e., those conforming to the {\scriptsize 
$PE^{\star}$} concept).

Since staging complete, \mi\ dialogs in this model does not require
verification of the order and size of utterances, the objective of the dialog
miner is to identify as much of the input dialog as possible that can be
specified through the {\scriptsize 
$PE^{\star}$} concept. In other words, it needs to identify
as much of the dialog as possible that can be handled by partial evaluation.
This process has been referred to as \textit{layering}~\cite{MIIMC}.  Moreover,
the objective of rewrite rules is not to reduce a complex dialog to one
expressed only through the
primitive concepts {\scriptsize $I$} or {\scriptsize $C$}. On the contrary,
rather we desire to express as much of the dialog as possible through the
{\scriptsize $PE^{\star}$} concept
to similarly improve the implementation.  For instance, it is
advantageous to express the specification 
{\scriptsize 
\{$\prec$a b$\succ$, $\prec$b
a$\succ$, $\prec$(a b)$\succ$\}} as
{\scriptsize $\frac{{PE}^{\star}}{\textrm{a b}}$} rather
than {\scriptsize 
$\frac{C}{\textrm{a b}} \cup \frac{C}{\textrm{b a}} \cup
\frac{I}{\textrm{a b}}$}.  Furthermore, to take advantage of all possible
opportunities to use partial evaluation, rewrite rules can be applied not only
to the original, pristine, script for a dialog, but also to the transformed,
reduced script remaining after every utterance.  For instance,
{\scriptsize 
$\mix
[\frac{SPE^{'}}{\frac{PE}{\textrm{a b}} \frac{PE}{\textrm{c d}}},
\mathtt{d}=\dots] = \frac{SPE^{'}}{\frac{PE}{\textrm{c}} \frac{PE}{\textrm{a
b}}} = \frac{SPE^{'}}{\frac{C}{\textrm{c}} \frac{PE}{\textrm{a b}}} =
\frac{C}{\textrm{c} \frac{PE}{\textrm{a b}}}$}, and
{\scriptsize 
$\mix [\frac{C}{\textrm{c}
\frac{PE}{\textrm{a b}}}, \mathtt{c}=\dots] = \frac{C}{\frac{PE}{\textrm{a b}}}
= \frac{PE}{\textrm{a b}}$}.  This can be thought of as relayering after
every utterance.  Unlike Figs.~\ref{fig:transexample} and~\ref{fig:transexample2},
where the script to be partial evaluated is the first argument to $\mix$, here,
for purposes of conserving space, we use a specification of the dialog in this
notation based on concepts from programming languages to represent the script to
be partially evaluated.

\subsection{Evaluation}

One method of quantifying control complexity or, more specifically,
anticipation of permissible orders and forms of utterances, is to count the
number of sub-expressions in a dialog specified
using this notation based on concepts from programming languages
that its stager must support.  In other words, the number of
sub-expressions required
to capture the requirements of a dialog is an evaluation
metric for the complexity of its stager. A complete, \mi\ dialog can be
captured by one expression.  If we remove only one of the thirteen episodes
from the 
{\scriptsize $\frac{{PE}^{\star}}{\textrm{size blend cream}}$} complete, \mi\
dialog, its requirements can no longer be captured by only
one expression. For
instance, a dialog where the 
{\scriptsize $\prec$(size blend cream)$\succ$} episode is absent
from the {\scriptsize 
$\frac{{PE}^{\star}}{\textrm{size blend cream}}$} specification cannot
be represented with less than five sub-expressions (i.e.,
{\scriptsize 
$\frac{SPE^{'}}{\textrm{size blend cream}} \cup \frac{SPE}{\textrm{size blend
cream}} \cup \frac{C}{\frac{I}{\textrm{size~blend}} \textrm{cream}} \cup
\frac{C}{\frac{I}{\textrm{blend~cream}} \textrm{size}} \cup
\frac{C}{\frac{I}{\textrm{size~cream}} \textrm{blend}})$}.  Therefore, in this
model, a stager for the prior dialog
is less complex in the control than one for the
latter.  While
{\scriptsize $\frac{{PE}^{\star}}{\textrm{size blend cream}}$} represents one
poset, note that there is not a one-to-one correspondence between posets and
expressions in this notation, based on concepts from
programming languages, that capture the requirements of
a dialog. For instance, while the previous dialog cannot be represented with a
union of less than five sub-expressions in this notation,
it can be represented by one poset.

When the specification for the dialog being staged cannot be captured by a
single expression (e.g., specifications c Table~\ref{tab:posets}
and 
{\scriptsize
$\frac{C}{\textrm{size} \frac{SPE}{\textrm{blend cream}}} \cup
\frac{C}{\textrm{blend} \frac{SPE}{\textrm{cream size}}} \cup
\frac{C}{\textrm{cream blend size}}$}) we
currently require one stager per expression.  Then a decision, based on the
user's first utterance, is required, if possible, to determine which stager to
invoke.  Staging dialogs in the $\Delta$ class that involve sub-dialogs
requires additional consideration.  A stager for each of these dialogs not only
needs to control how partial evaluation is invoked to support the individual
concepts, but also needs to coordinate the order in which it jumps into and
returns from sub-dialogs.

\begin{lstlisting}[float,basicstyle=\tiny,frame=tblr,
stepnumber=0,caption={Generalized stager algorithm simplified for purposes
of presentation.},label=lst:stager]
(generate-pristine-sub-scripts-from-expressions)
(let* ((origin-script (file->list script-file-name))
       (origin-counter 1)
       (loop-key
          (call/cc
             (lambda (k) 
                (list k origin-script origin-counter))))
       (loop (car loop-key))
       (script (cadr loop-key))
       (counter (caddr loop-key)))
  
  (prompt-for-input)
  
  (if (valid-input?)
     (let ((new-script (partially-eval-script-with-utterance)))
        (if (complete-evaluation?)
           (eval new-script)
           (k (list k new-script counter+1))))
     (k (list k script counter)))) ; invalid input
\end{lstlisting}

\subsection{Practical Considerations}

We made a few practical considerations in our system implementation.  For
instance, since we are only using the constant propagation capability of
partial evaluation, we implemented our own language processing function that
propagates the supplied arguments throughout the body of a script, rather than
using an off-the-shelf partial evaluator.  We also capture first-class
continuations through the \texttt{call/cc} facility in Scheme to restart a
stager after each progressive partial evaluation.  We generate a loop key,
which contains a continuation, a script, and an occurrence counter, with the
initial, pristine script.  A stager then prompts for and accepts an utterance
from the user, validates the responses in the utterance and, if valid, marshals
it into a set of parameter-value pairs, and partially evaluates a script with
that set of static parameter assignments.  Finally, the stager generates a new
loop key with the specialized script and uses the continuation to jump back to
the beginning of the stager. The generalized stager algorithm shown in
Listing~\ref{lst:stager} illustrates this process.  We also incorporate a
history-records object, which stores the loop key of each stage, to support
\texttt{redo} and \texttt{undo} operations.
The functional paradigm of programming enabled us to support these finer
touches in a simple and clean manner.

\subsection{Stager Generation}

\begin{figure}
\centering
\begin{tabular}{c}
\includegraphics[scale=0.40]{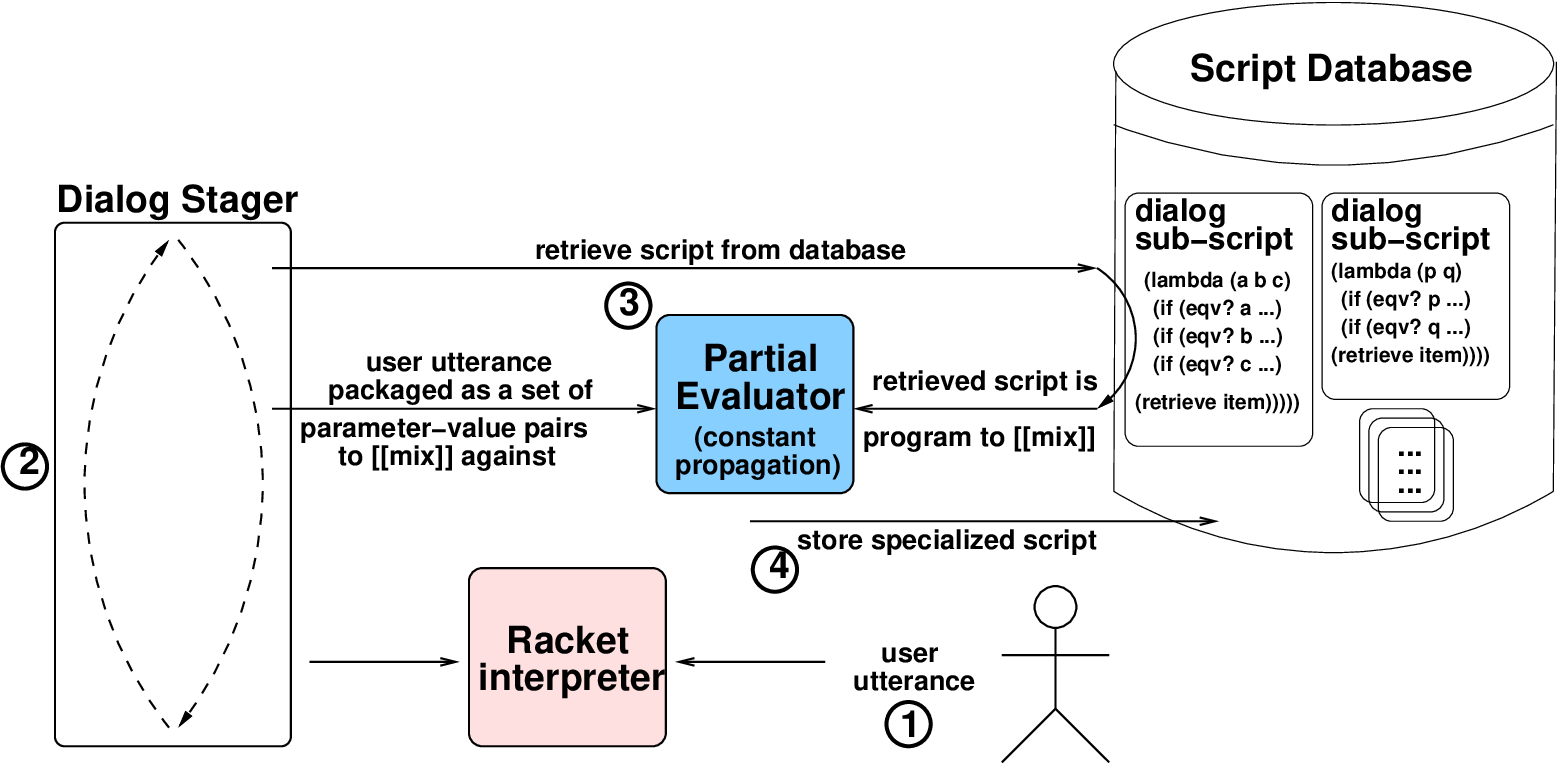}\\
\end{tabular}
\caption{Stager system design and execution. Numbers indicate
flow of progression and this progression repeats until the dialog is complete.}
\label{fig:arch}
\end{figure}

From an expression, in notation based on concepts from programming languages,
output by the dialog miner, we automatically generate a stager to realize
and execute the
dialog (see transition from the center to the right of Fig.~\ref{fig:overview}).
Since Scheme is a homoiconic language, it is suitable for generating a stager
in Scheme.
Given a dialog specified in the notation, based on concepts from programming
languages, presented here, it automatically generates a stager, and the
necessary scripts.
The size of the stager it generates is dependent on the complexity of the
dialog to be staged.
The generated stagers must be executed with \textit{DrRacket} (version 6.0) with
the language set to `Determine language from source'.
Fig.~\ref{fig:arch} provides an overview of the execution of the stagers we
generate (i.e., the rightmost side of Fig.~\ref{fig:overview}).  Our dialog
modeling toolkit, including the dialog miner and stager generator, is available
at \url{http://academic.udayton.edu/SaverioPerugini/dialogtoolkit.zip}.

\section{Discussion}
\label{sec:discussion}

The advent and increased use of virtual, immersive environments in
cyberlearning~\cite{hufpost}, gaming, and smart phones apps provides a new
landscape and opportunity to research models for designing and implementing
flexible human-computer dialogs.  Cyberlearning activities and apps whose
success relies on flexible, \mi\ dialog can benefit from a model for designing
and implementing dialogs in a more systematic and simplified way.  Our project
is particularly worthwhile and applicable in the context of improving the
implementation of dialogs in these cyberlearning environments because through a
mixed-initiative motif, our research supports and fosters enhanced
teacher-student collaboration by blurring the boundaries between teacher and
student.  Moreover, our model for \mii\ has the potential to be widely adopted
due to the growing number of cyberlearning applications~\cite{hufpost} and
smart phone apps and users of these systems.  Thus, while this research project
takes a non-traditional approach to dialog modeling and implementation, we are
optimistic that it can have an impact on rhe software development process for
virtual/cyberlearning environments, gaming, and smart phone apps, where
flexibility in human-computer dialog is important.

Prior research projects have approached intelligent information system design
from the perspective of (functional) programming
languages~\cite{automaticallyRestrWPs,FPLweb,uic,continuationsWebServers}.
However, only a few research projects have sought to marry human-computer
dialogs with concepts from programming
languages~\cite{vt-staging,FPLweb,PerezDissertation,MIIMC}.  Due to the
conceptual metaphors between natural languages and programming languages,
viewing human-computer dialog modeling and implementation from the perspective
of programming languages suggests a natural, yet under-explored, approach to
dialog representation and reasoning.  Thus, we also expect our work to generate
discussion in the automated software engineering community.

\subsection{Contributions}

We modified and improved a model for specifying and staging \mi\
dialogs~\cite{vt-staging,WWW2004,MIIMC}.  The primary theme of this approach is
to identify as many embedded complete, \mi\ sub-dialogs in a specification
since they can be advantageously handled by partial evaluation.  Identifying
these sub-dialogs permits one to plan for only one episode in a script and
judiciously apply partial evaluation to it to realize all possible variations
of that episode.  This non-traditional use of partial evaluation helps meet the
specification of a dialog without having to explicitly hardcode all supported
episodes into the control flow of the implementation.  We generalized the
activity of automatically building a stager and generate stagers for a variety
of unsolicited reporting, \mi\ dialogs.  While ``[c]reating an actual dialog
system involves a very intensive programming effort''~\cite{evalMII}, our
toolkit automates the construction of a \textit{stager} from a specification of
a \mi\ dialog in our notation.  Three dialog models, namely transition
networks, context-free grammars, and events, have been identified
in~\cite{3dialogModels}.  We feel that we are at a vanguard of a fourth dialog
model.  We have extended~\cite{vt-staging,WWW2004,MIIMC} in multiple ways.  We
summarize our contributions as: we

\begin{itemize}

\item recognize the need to support multiple orders of responses independent of
multiple responses per utterance and to bring more structure to the space of
unsolicited reported, \mi\ dialogs. To do so we enriched and augmented a
notation, based on concepts from programming languages, for specifying dialogs
by adding and modifying concepts and their mnemonics;

\item introduce a dialog mining component, including
the layering essential to, and deemed critical 
in~\cite{MIIMC} for, implementing dialogs containing
nested sub-dialogs.  Ramakrishnan, Capra, and
P{\'{e}}rez-Qui{\~{n}}ones note that developing an initial,
optimal representation of a dialog is an open research issue
in~\cite{MIIMC}:

\begin{quote} An interesting research issue is: [g]iven (i) a set of
interaction sequences [(referred to as episodes here)], and (ii) addressable
information (such as arguments and slot variables), determine (iii) the
smallest program so that every interaction sequence can be staged \dots. [T]his
requires algorithms to automatically decompose and `layer' interaction
sequences into those that are best addressed [by an] interpreter and those that
can benefit from representation and specialization by [a] partial
evaluator~\cite{MIIMC}.  \end{quote}

\noindent
Our dialog miner addresses this issue.

\item automatically generate stagers for a variety of unsolicited reporting,
\mi\ dialogs, including those involving sub-dialogs;

\item encompassed all of above in a free, downloadable dialog modeling toolkit.

\end{itemize}

\subsection{Future Work}

We intend to widen the scope of the unsolicited reporting, \mi\ dialogs that
can be accommodated (i.e., specified and staged) in this model.  For
instance, we intend to enhance our mining and layering algorithms so that we can
recognize and stage dialogs, involving more than one sub-dialog, specified with
an {\scriptsize 
$SPE^{'}$} in the numerator such as dialog d in Table~\ref{tab:posets}.  Since
Scheme supports first-class and higher-order functions, we intend to explore
partially evaluating scripts that, unlike that shown in
Fig.~\ref{fig:transexample},
accept functions representing scripts for sub-dialogs as parameters rather than
individual responses.  Moreover, we are working on algorithms
to deal with dialogs where the episodes therein cannot be represented by a
single poset (e.g., dialog c in Table~\ref{tab:posets}).
We have identified specific examples where a dialog cannot be specified with
less than $y$ posets, yet can be staged using $x$ scripts, where $x < y$.  We
intend to study such cases for insight into solving this problem in general.

Beyond these issues, we intend to lift additional restrictions on the space of
unsolicited reporting, \mi\ dialogs to further expand the space of dialogs on
which we work.  For example, not all dialogs have a consistent number of
questions across all episodes.  More generally, some dialogs have dependencies
between responses as identified in~\cite{JASIST2007}, in a slightly different
context.  In the dialogs we have presented in this article, due to domain
semantics, the answers to the questions posed are completely independent of
each other.  In other words, any answer to any question does not disqualify any
of the answers to any of the other questions.  However, in other domains such
complete independence may not exist.  For example, the 2014 Honda Civic Hybrid
is not available with a manual transmission and, therefore, there is no need to
prompt for transmission type once Honda and Civic Hybrid are specified for make
and model, respectively.  Therefore, we must study how to programmatically
represent dependencies between the responses in a dialog.  We plan to explore a
variety of options to deal with dependencies.
In covering a richer assortment of unsolicited reporting, \mi\ dialogs, we
intend to evolve the dialog notation into a domain-specific language, and the
toolkit into a dialog prototyping tool, which dialog designers can use to
explore and test a variety of unsolicited reporting, \mi\ dialogs.

Nonetheless, communicating independent responses in a variety of orders and
combinations are practical aspects of common dialogs, as demonstrated in
Table~\ref{tab:pracspace}.  Thus, we feel this approach and project is worthwhile
and potentially applicable in the context of improving the implementation of
dialogs within smart phone apps.
We feel that both the philosophical and conceptual
connections between natural and programming
languages~\cite{lingusticsideeffects} suggest that additional concepts from
programming languages,
such as reflection and first-class continuations, may find a place in this
model.

Dialog is essential to providing a rich form of human-computer
interaction. This project seeks to establish a sound and well-evaluated
computational model for mixed-initiative dialog. 
We envisage the long-term practical significance and broader impacts of this
work involving the incorporation of stagers based on partial evaluation and
rewrite rules into cyberlearning environments,
smart phones, airport kiosks, \textsc{itm} machines, and
interactive, voice-responses systems, since the ubiquity of these platforms in
service-oriented domains, such as education, health care, banking,
and travel provide a fertile landscape for the use of our model for
\mii.

\section*{Acknowledgments}

The research was supported in part by grants from the Ohio Board of Regents and
support from the University of Dayton College of Arts and Sciences.  We thank
John Cresencia, Shuangyang Yang, and Brandon Williams at the University of
Dayton for assisting in the implementation of the dialog modeling toolkit, and
for helpful discussions and insight.

\bibliographystyle{plain}
\bibliography{staging}

\begin{thebibliography}{10}

\bibitem{MII-UR}
J.F. Allen.
\newblock Mixed-initiative interaction.
\newblock {\em IEEE Intelligent Systems}, 14(5):14--16, 1999.

\bibitem{termrewriting}
F.~Baader and T.~Nipkow.
\newblock {\em Term Rewriting and All That}.
\newblock Cambridge University Press, Cambridge, UK, 1999.

\bibitem{vt-staging}
R.~Capra, M.~Narayan, S.~Perugini, N.~Ramakrishnan, and M.A.
  P{\'{e}}rez-Qui{\~{n}}ones.
\newblock The staging transformation approach to mixing initiative.
\newblock In G.~Tecuci, editor, {\em Working Notes of the IJCAI 2003 Workshop
  on Mixed-Initiative Intelligent Systems}, pages 23--29, Menlo Park, CA, 2003.
  AAAI/MIT Press.

\bibitem{hufpost}
A.~Dubrow.
\newblock Seven cyberlearning technologies transforming education.
\newblock \textit{Huffington Post}, 6 April 2015. Available:
  \url{http://www.huffingtonpost.com/aaron-dubrow/7-cyberlearning-technolog_b_6988976.html}
  [Last accessed: 18 November 2015].

\bibitem{EOPL3}
D.P. Friedman and M.~Wand.
\newblock {\em Essentials of Programming Languages}.
\newblock MIT Press, Cambridge, MA, third edition, 2008.

\bibitem{automaticallyRestrWPs}
P.~Graunke, R.~Findler, S.~Krishnamurthi, and M.~Felleisen.
\newblock Automatically restructuring programs for the web.
\newblock In {\em Proceedings of the Sixteenth IEEE International Conference on
  Automated Software Engineering (ASE)}, pages 211--222, 2001.

\bibitem{3dialogModels}
M.~Green.
\newblock {A Survey of Three Dialogue Models}.
\newblock {\em ACM Transactions on Graphics}, 5(3):244--275, July 1986.

\bibitem{evalMII}
C.I. Guinn.
\newblock Evaluating mixed-initiative dialog.
\newblock {\em IEEE Intelligent Systems}, 14(5):21--23, 1999.

\bibitem{introPartialEvaluation}
N.D. Jones.
\newblock {An Introduction to Partial Evaluation}.
\newblock {\em ACM Computing Surveys}, 28(3):480--503, September 1996.

\bibitem{stagingMLHOSC}
Y.D. Liu, C.~Skalka, and S.F. Smith.
\newblock {Type-specialized Staged Programming with Process Separation}.
\newblock {\em Higher-Order and Symbolic Computation}, pages 1--45, September
  2012.

\bibitem{FPLweb}
S.N. Malkov.
\newblock {Customizing a Functional Programming Language for Web Development}.
\newblock {\em {Computer Languages, Systems and Structures}}, 36(4):345--351,
  2010.

\bibitem{WWW2004}
M.~Narayan, C.~Williams, S.~Perugini, and N.~Ramakrishnan.
\newblock Staging transformations for multimodal web interaction management.
\newblock In M.~Najork and C.E. Wills, editors, {\em Proceedings of the
  Thirteenth International {ACM} World Wide Web Conference ({WWW})}, pages
  212--223, New York, NY, 2004. {ACM} Press.

\bibitem{PerezDissertation}
M.A. P{\'{e}}rez-Qui{\~{n}}ones.
\newblock {\em {Conversational Collaboration in User-initiated Interruption and
  Cancellation Requests}}.
\newblock {Ph.D.} dissertation, The George Washington University, 1996.

\bibitem{JASIST2007}
S.~Perugini and N.~Ramakrishnan.
\newblock Mining web functional dependencies for flexible information access.
\newblock {\em Journal of the American Society for Information Science and
  Technology ({JASIST})}, 58(12):1805--1819, 2007.

\bibitem{uic}
D.~Quan, D.~Huynh, D.R. Karger, and R.~Miller.
\newblock {User Interface Continuations}.
\newblock In {\em Proceedings of the Sixteenth Annual ACM Symposium on User
  Interface Software and Technology (UIST)}, pages 145--148, New York, NY,
  November 2003. ACM Press.

\bibitem{continuationsWebServers}
C.~Queinnec.
\newblock The influence of browsers on evaluators or, continuations to program
  web servers.
\newblock In {\em Proceedings of the Fifth ACM SIGPLAN International Conference
  on Functional Programming (ICFP)}, pages 23--33, New York, NY, 2000. ACM
  Press.
\newblock Also appears in \textit{ACM SIGPLAN Notices}, 35(9), 2000.

\bibitem{MIIMC}
N.~Ramakrishnan, R.~Capra, and M.A. P{\'{e}}rez-Qui{\~{n}}ones.
\newblock {Mixed-Initiative Interaction = Mixed Computation}.
\newblock In P.~Thiemann, editor, {\em {Proceedings of the ACM SIGPLAN Workshop
  on Partial Evaluation and Semantics-Based Program Manipulation (PEPM)}},
  pages 119--130, New York, NY, 2002. ACM Press.
\newblock Also appears in \textit{ACM SIGPLAN Notices}, 37(3), 2002.

\bibitem{lingusticsideeffects}
C-C. Shan.
\newblock {Linguistic Side Effects}.
\newblock In C.~Barker and P.~Jacobson, editors, {\em Direct Compositionality},
  pages 132--–163. Oxford University Press, 2007.

\end{thebibliography}

\end{document}